\documentclass[apj]{emulateapj}
\usepackage{epsfig}
\usepackage{natbib}
\usepackage{lscape}
\usepackage{rotating}
\usepackage{graphicx}

\newcommand{\ha}{\textrm{H}\ensuremath{\alpha}}
\newcommand{\hb}{\textrm{H}\ensuremath{\beta}}

\newcommand{\oi}{[\textrm{O}~\textsc{i}]}

\newcommand{\oiiilam}{[\textrm{O}~\textsc{iii}]~\ensuremath{\lambda5007}}
\newcommand{\niilam}{[\textrm{N}~\textsc{ii}]~\ensuremath{\lambda6584}} 
\newcommand{\oiilam}{[\textrm{O}~\textsc{iii}]~\ensuremath{\lambda3727}}
\newcommand{\Lphcn}{$L^{\prime}_{\rm{HCN}}$}
\newcommand{\LphcnOne}{$L^{\prime}_{\rm{HCN(1-0)}}$}
\newcommand{\LphcnThr}{$L^{\prime}_{\rm{HCN(3-2)}}$}

\newcommand{\LphcoOne}{$L^{\prime}_{\rm{HCO^{+}(1-0)}}$}
\newcommand{\LphcoThr}{$L^{\prime}_{\rm{HCO^{+}(3-2)}}$}
\newcommand{\Lpco}{$L^{\prime}_{\rm{CO}}$}
\newcommand{\LpcoOne}{$L^{\prime}_{\rm{CO(1-0)}}$}

\newcommand{\coOne}{$\rm{CO(1-0)}$}
\newcommand{\coThr}{$\rm{CO(3-2)}$}
\newcommand{\hcnOne}{$\rm {HCN(1-0)}$}
\newcommand{\hcnThr}{$\rm{HCN(3-2)}$}
\newcommand{\hcoOne}{$\rm{HCO^{+}(1-0)}$}
\newcommand{\hcoThr}{$\rm{HCO^{+}(3-2)}$}
\newcommand{\lmol}{\mbox{$L_{\rm mol}$}}

\slugcomment{To Appear in the ApJ}

\shorttitle{Enhanced Gas Density in ULIRGs}
\shortauthors{Juneau et al.}

\begin{document}

\title{Enhanced Dense Gas Fraction in Ultra-Luminous Infrared Galaxies} 

\author{S. Juneau\altaffilmark{1}, D. T. Narayanan\altaffilmark{2,3}, 
  J. Moustakas\altaffilmark{4,5}, Y. L. Shirley\altaffilmark{1}, 
  R. S. Bussmann\altaffilmark{1}, R. C. Kennicutt Jr.\altaffilmark{6},
  \& P. A. Vanden Bout\altaffilmark{7}}
\altaffiltext{1}{Steward Observatory,
  Department of Astronomy, University of Arizona, 933 N. Cherry Ave.,
  Tucson, AZ 85721; sjuneau@as.arizona.edu;
  yshirley@as.arizona.edu; rsbussmann@as.arizona.edu}
\altaffiltext{2}{Harvard-Smithsonian Center for Astrophysics, 60
  Garden St. MS 51, Cambridge, MA, 02138, dnarayanan@cfa.harvard.edu}
\altaffiltext{3}{CfA Fellow} 
\altaffiltext{4}{Department of Physics, New York University, 4
  Washington Place, New York, NY 10003}
\altaffiltext{5}{Center for Astrophysics and Space Sciences, 
  University of California, San Diego, 9500 Gilman Drive, La Jolla, 
  CA 92093-0424; jmoustakas@ucsd.edu}
\altaffiltext{6}{Institute of Astronomy,
  University of Cambridge, Madingley Road, Cambridge CB3 0HA, UK;
  robk@ast.cam.ac.uk}
\altaffiltext{7}{National Radio Astronomy Observatory, 520 Edgemont 
  Road, Charlottesville, VA 22903; pvandenb@nrao.edu}

\begin{abstract} 

We present a detailed analysis of the relation between infrared 
luminosity and molecular line luminosity, for a variety of molecular 
transitions, using a sample of 34 nearby galaxies spanning a broad 
range of infrared luminosities ($10^{10} < L_{IR} < 10^{12.5} L_{\sun}$).
We show that the power-law index of the relation is sensitive to 
the critical density of the molecular gas tracer used, and that the 
dominant driver in observed molecular line ratios in galaxies is 
the gas density.  As most nearby ultraluminous infrared galaxies (ULIRGs) 
exhibit strong signatures of active galactic nuclei (AGN) in their center, 
we revisit previous claims questioning the reliability of HCN as a probe 
of the dense gas responsible for star formation in the presence of AGN.  
We find that the enhanced HCN(1-0)/CO(1-0) luminosity ratio observed in 
ULIRGs can be successfully reproduced using numerical models with fixed 
chemical abundances and without AGN-induced chemistry effects.  We extend 
this analysis to a total of ten molecular line ratios by combining the 
following transitions: \coOne, \hcoOne, \hcoThr, \hcnOne, and \hcnThr.
Our results suggest that AGNs reside in systems with 
higher dense gas fraction, and that chemistry or other effects associated 
with their hard radiation field may not dominate (NGC~1068 is one exception).  
Galaxy merger could be the underlying cause of increased dense gas 
fraction and the evolutionary stage of such mergers may be another 
determinant of the HCN/CO luminosity ratio.

\end{abstract}

\keywords{galaxies: evolution --- galaxies:fundamental parameters and
  ISM --- infrared: galaxies --- radio lines: galaxies ---
  submillimeter}

\section{Introduction}
\label{sec:intro}

Over the past 50 years, there have been a number of studies relating
galaxy star formation rate (SFR) and the amount of molecular gas
available.  \citet{1959ApJ...129..243S} proposed a power-law
relationship between star formation rate volume density and molecular
gas volume density.  \citet{1998ApJ...498..541K} 
\citep[also see][]{1989ApJ...344..685K} framed the problem
in terms of surface densities, yielding the Kennicutt-Schmidt
relation: $\Sigma_{SFR} \propto \Sigma_{gas}^{1.4 \pm 0.15}$.

While a power-law index around 1.4 or 1.5 holds for the molecular gas traced
by \coOne, a power-law index of unity was found when relating SFR to
the molecular gas traced by \hcnOne\ \citep[hereafter
  GS04b]{1992ApJ...387L..55S,  2004ApJ...606..271G}.  This difference
in index is interpreted as a consequence of the different critical
densities\footnote{$n_{crit} \equiv A_{ul}/\Gamma_{ul}$, where $A_{ul}$ 
is the Eintein A coefficient and $\Gamma_{ul}$ is the collision rate 
for a transition from upper (u) to lower (l) energy level.} of the 
molecular tracers used.  Having a lower critical density, \coOne\ 
traces the global molecular gas content, some of which may not be 
involved in the star formation process, whereas \hcnOne\ traces 
denser molecular gas ($n > 10^{4}~cm^{-3}$) more closely linked to 
star formation. Since the \citet{1992ApJ...387L..55S} and 
\citet[hereafter GS04a]{2004ApJS..152...63G} observations, 
several interpretations regarding the linear $L_{\rm
  IR}$-\hcnOne \ relation in galaxies have been put forth by both
observational arguments and numerical models.

For example, by extending the linear relationship between the total
infrared luminosity, $L_{\rm IR} [\equiv L$(8-1100$\mu$m)], and 
\hcnOne \ to Galactic cloud cores, \citet{2005ApJ...635L.173W} 
added to the original GS04b interpretation by framing the 
relationship in terms of individual star-forming dense-gas ``units''. 
As one increases the number of dense star-forming clumps in a galaxy, 
$L_{\rm IR}$ and \hcnOne \ both increase in lock-step.  In this view, 
the only difference between an extreme starburst galaxy and Galactic 
star-forming region is the number of dense star-forming units emitting 
HCN and infrared emission.

Recent observations may challenge the existence of 
the tight, linear correlation between $L_{\rm IR}$ and \hcnOne \ 
luminosity\footnote{We use the notation L' for molecular line 
luminosities as they are expressed in K km s$^{-1}$ pc$^2$, whereas  
other luminosities are expressed in L$_{\sun}$.} reported by GS04b.  
For instance, \citet[hereafter GC08]{2008A&A...479..703G} obtained
new HCN and HCO$^{+}$ measurements for 17 luminous and ultra-luminous 
infrared galaxies (LIRGs: $10^{11} < L_{IR} < 10^{12}~L_{\sun}$ and 
ULIRGs: $L_{IR} > 10^{12}~L_{\sun}$).  They found
conflicting \hcnOne\ measurements for several galaxies that overlap
with the \citet{1992ApJ...387L..55S} sample.
They attribute the differences to calibration
errors in the older survey, and proceed to compare their revised
$L_{IR}-$\Lphcn\ relation with the original one.  They find a larger
and more significant increase of $L_{IR}-$\Lphcn\ with $L_{IR}$,
meaning that the $L_{IR}-$\Lphcn\ power-law would be steeper than an
index of unity.  These authors interpreted the higher $L_{\rm
  IR}$/\Lphcn\ ratio as an enhanced star formation efficiency in
ULIRGs.

A large fraction of nearby LIRGs and ULIRGs appear to be mergers of gas-rich 
progenitor galaxies and therefore have a high concentration of dense 
molecular gas in their center which feeds strong starburst and AGN
activity \citep[see][and references therein]{1996ARA&A..34..749S}.
It has been noted that sources with brighter IR luminosities are both more 
likely to host an AGN \citep[e.g.][]{1995ApJS...98..171V, 1998ApJ...505L.103L, 1999ApJ...522..113V}, 
as well as show higher \hcnOne/\coOne\ molecular line luminosity ratios 
\citep[GS04b,][]{2006ApJ...640L.135G}. It is tempting to associate 
the increased \hcnOne/\coOne \ luminosity ratio with the presence of AGN. 
Indeed, \citet{2008A&A...479..703G} report evidence for an enhanced 
abundance of HCN in ULIRGs, and caution that care must be taken when 
converting HCN line luminosities into dense gas masses.

Moreover, some theoretical work suggests that 
AGN-induced processes could potentially increase the abundance of
HCN \citep[e.g.][]{2006ApJ...646L..37L}, causing the observed $L_{\rm
  IR}$-\Lphcn \ relation to have a shallower slope than the $L_{\rm
  IR}$-\Lpco \ relation. In this picture, X-ray emission resulting
from AGN buried in ULIRGs may enhance the abundance of HCN molecules
via an increased availability of free electrons that facilitates 
combination with ions (e.g., HCNH$^+$ + e$^-$ $\rightarrow$ HCN + H). 
Some high-resolution HCN observations of nearby galaxies 
have shown stronger emission of \hcnOne\ relative to \coOne\ 
in the center of a few LIRGs and ULIRGs known to host an AGN 
\citep[e.g.][]{2006ApJ...640L.135G,2007A&A...468L..63K}. These 
authors have interpreted this higher molecular line ratio in terms of 
an increased abundance of HCN.  Others interpret high molecular line
ratios in terms of the mid-plane pressure, in the sense that larger 
ambient pressure could trap or maintain dense gas \citep{2008ApJ...673..183L, 2006ApJ...650..933B}.

In recent years, new theoretical models suggest a
different physical explanation for the observed $L_{\rm IR}$-molecular
line luminosity relationships.  Specifically, models that couple
non-local thermodynamic equilibrium (LTE) radiative transfer
calculations with models of star-forming giant molecular clouds 
(GMCs) \citep{2007ApJ...669.289K} and hydrodynamic simulations of disk
galaxies and galaxy mergers \citep{2008ApJ...684..996N} find that the
driving force controlling the SFR-\lmol \ relation in galaxies is the
fraction of thermalized molecular gas at a given molecular
transition. In the \citet{2007ApJ...669.289K} and
\citet{2008ApJ...684..996N} picture, the SFR is controlled by the
relation SFR$\propto n^{1.5}$, where $n$ is the number density of
molecular hydrogen.  When the observed molecular line traces the 
bulk of the molecular gas in the galaxy (e.g. \coOne), the SFR-\lmol
\ power-law index (hereafter ``slope'' as this relationship is
typically considered in log-log space) is equivalent to the
Kennicutt-Schmidt index (i.e. close to 1.5). On the other hand, higher 
critical density molecular line tracers (e.g. \hcnOne) trace an increasingly 
smaller fraction of the gas in a galaxy, and so the index in the observed 
SFR-\lmol\ power-law relation decreases. These theories were tested by
\citet{2008ApJ...681L..73B}, who observed \hcnThr \ --- a higher critical
density tracer than \hcnOne --- in the GS04a,b sample of galaxies, and
found a SFR-\hcnThr \ index less than unity ($0.72 \pm 0.08$).

In this paper, we expand on the Bussmann et al. work by studying the 
relationship between five molecular line tracers and infrared luminosity. 
In particular, we emphasize the importance of considering the critical 
density of each molecular line tracer.  
We present support that \hcnOne\ is a valid probe of dense molecular gas and  
we revisit the interpretation of the enhanced \hcnOne/\coOne\ luminosity 
ratio in high IR luminosity systems. Our results suggest that this feature is
primarily driven by the increased dense gas fraction in galaxies undergoing a
merger and/or a strong episode of starburst activity; we do not require X-ray
driven chemical abundance effects to explain the observed molecular line ratios.
Overall, we build a coherent picture where the governing parameter is the 
density distribution of the molecular gas.

We describe our sample of 34 galaxies in \S\ref{sec:method}, along with 
the measurements that we use.  Section~\ref{sec:results} contains our 
results regarding power-law relationships between infrared luminosity and 
various molecular line luminosities (\S\ref{sec:lum}), as well as the study of 
molecular line ratios as a function of infrared luminosity (\S\ref{sec:lineratios}).
The interpretation is partially based on the theoretical models of 
\citet{2008ApJ...684..996N}, which we apply to this study in \S\ref{sec:model}.  
We include a brief discussion of chemistry effects in \S\ref{sec:Chem} and 
present our conclusions in \S\ref{sec:conclusions}.  Throughout this paper 
we assume a flat $\Lambda$CDM cosmological model with ($\Omega_{M}, 
\Omega_{\Lambda}$, H$_0$) = (0.3, 0.7, 70~km s$^{-1}$ Mpc$^{-1}$).

\section{Methods}\label{sec:method}

\subsection{Sample}
\label{sec:sample}

Our primary sample consists of 29 nearby galaxies with
\hcnOne\ observations \citep{2004ApJ...606..271G} as well as
integrated optical spectroscopy \citep[hereafter
  MK06]{2006ApJS..164...81M}.  We obtained follow-up
\hcnThr\ observations for 22 of these 29 galaxies (10 detections and
12 upper limits) using the 10~m Heinrich Hertz Submillimeter Telescope
\citep[see][hereafter B08 for more detail on the \hcnThr\ observations]{2008ApJ...681L..73B}.  
In order to expand our sample and include additional transitions 
spanning a range of critical densities and excitation 
states, we incorporate the sample from \citet[hereafter
  GC08]{2008A&A...479..703G}.  These authors include observations of
the \hcnOne, \hcnThr, \hcoOne, \hcoThr\ transitions for 17 galaxies,
of which 12 overlap with our original sample.  Thus, the combined
sample comprises 34 galaxies.  We note a distinction between our
primary sample and that of GC08.  While the GC08 sample spans the high
infrared luminosities ($L_{IR} = 10^{11.3} - 10^{12.5} L_{\sun}$), our
sample extends the range down to $L_{IR} \sim 10^{10}~L_{\sun}$.

Molecular line and (far-)infrared luminosities are tabulated in
Table~\ref{tab:data}.  When combining our primary sample and GC08
sample, we convert all the luminosities to common cosmological
parameters and luminosity distances.  The \coOne\ and
\hcnOne\ measurements were taken from GS04a,b.  When available, we
substitute updated \hcnOne\ luminosities from GC08.  These authors
published new \hcnOne\ luminosities for their 17 galaxies and found
significantly different values for galaxies overlapping with the 
\citet{1992ApJ...387L..55S} subsample.  These galaxies were also included 
in the GS04b compilation and the new values from GC08 are 
approximately a factor of two lower in several cases.  GC08 attributed
these discrepancies to observational errors, namely the calibration of
the receiver used on the 30-m IRAM telescope during the earlier
observations.

\LphcnThr\ values were obtained from B08 (see Appendix~\ref{sec:append})
 and GC08.  Among eight galaxies in common, five have consistent 
values within the uncertainty, in which case we use the average 
(or the maximum upper limit in the case of two upper limits).  
We adopt the GC08 measurements for two non-detections in B08 (Mrk~231 
\& IRAS~17208-0014), as well as for NGC~6240, which had a discrepant
measurement.  This galaxy has a very broad \hcnThr\ emission line,
so in this case we prefer to use the data from GC08 because of their
significantly wider bandwidth.

A few galaxies in our sample were also observed by \citet{2008ApJ...677..262K}.  
These authors report brighter \hcnThr\ for five overlapping 
galaxies, most notably for the most IR-luminous galaxies (e.g. Mrk 231).  
To maintain homogeneity in our selected set of observations, and to 
facilitate the comparison with the work of GC08, we restrict our 
analysis to the \hcnThr\ luminosities from B08 and GC08.  However, 
we anticipate that including the few higher values from 
\citet{2008ApJ...677..262K} would either strenghten the trends 
presented in \S\ref{sec:results} or leave them unchanged.

We use the \LphcoOne\ and \LphcoThr\ measurements published in GC08
with a conversion for the luminosity distances adopted here.  On
average, this distance conversion changes molecular line luminosities
by 0.055~dex (min=-0.073 and max=0.16), which is less than the average
uncertainty on these quantities.

Observing different molecular transitions with a variety of telescopes
leads to varying beam sizes.  Since most sources in the samples we use
in this study have not been mapped, it is important to consider
aperture affects associated with the use of varying beam sizes.  First,
two galaxies have a lower limit in \hcnOne\ because they were not
mapped even though they are nearby (NGC~660 and NGC~2903, see
GS04a).  For the HCN(3-2) observations in B08, the beam size of ~30"
is sufficient to cover the central kpc for galaxies beyond 7 Mpc.  We
expect to detect most of the HCN(3-2) emission in these cases since it
is the dense nuclear regions of galaxies that are responsible for the
majority of the emission from molecular transitions with high critical
densities for excitation.  Meanwhile, targets selected for study in
GC08 all lie at distances $>$60 Mpc and therefore we are assured of 
sampling the full extent of the high density molecular emission.

\subsection{Optical AGN Classification}
\label{sec:BPT}

\begin{figure}
\epsscale{0.9}
\plotone{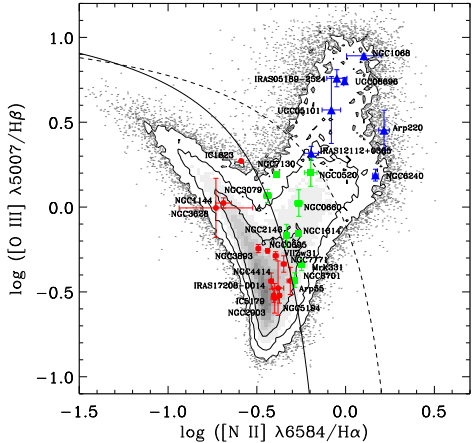}
\caption{Emission-line diagnostic diagram
  indicating the position of normal star-forming galaxies ({\em below
    and to the left of the solid curve}), AGNs ({\em above and to the
    right of the dashed curve}), and galaxies with an admixture of
  star formation and AGN activity ({\em between the solid and dashed
    curves}).  The individual galaxies in our sample are labeled and
  plotted using color symbols with error bars, with the exception
  of five galaxies mentioned in the text.
  The different symbols represent the optical spectral type: star-forming
  (SF) ({\em red circles}); AGN ({\em blue triangles}); and SF/AGN
  ({\em green squares}).  For reference, the gray-scale and contours
  (enclosing $52\%$, $84\%$, and $97\%$ of the points) show the locus
  of emission-line galaxies in the SDSS.  The solid and dashed curves
  are adapted from \citet{2003MNRAS.346.1055K} and
  \citet{2001ApJS..132...37K}, respectively.
  (A color version of this figure is available in the online journal.)}
\label{fig:BPT}
\end{figure}

The nature of the powering source of galaxies in our sample (star
formation, AGN, or hybrid i.e. hosting both star formation and AGN
activity) is determined using the optical diagnostic diagram known as
the {\it BPT diagram} \citep{1981PASP...93....5B, 1987ApJS...63..295V,
  1989agna.book.....O}.  The optical emission-line ratios used in this
diagram (\oiiilam/\hb\ and \niilam/\ha) probe a combination of the
oxygen abundance and ionization parameter of the interstellar medium
(ISM) present in these galaxies.  This provides us with an indirect
signature of the source of ionizing radiation (young stars versus
AGN).  As shown in Figure~\ref{fig:BPT}, the Sloan Digital Sky Survey 
\citep[SDSS,][]{2000AJ....120.1579Y} galaxies (gray area and contours) 
show a tight excitation 
sequence of star-forming galaxies (below the solid line from 
\citet{2003MNRAS.346.1055K}), as well as a plume of galaxies that 
are either hybrid (between the solid and dashed lines) or AGN 
(above the dashed line adapted from \citet{2001ApJS..132...37K}).  
The SDSS emission-line measurements are taken from the MPA/JHU SDSS team
website\footnote{http://www.mpa-garching.mpg.de/SDSS/DR7} 
\citep{2004MNRAS.351.1151B, 2004ApJ...613..898T} based on observations of 
galaxies contained in the SDSS data release~7 \citep[DR7;][]{2009ApJS..182..543A}.

Galaxies from our combined sample are
overlaid with colored symbols.  Among 34 galaxies, 13 are classified
as star-forming (SF, red circles), 10 galaxies are classified as
hybrid (SF/AGN, green squares) and 11 galaxies show a strong signature 
of AGN (blue triangles).  These numbers include five galaxies which
are classified but not shown on the diagram. Two galaxies (Mrk~231 and
NGC~7469) have obvious broad emission lines in their optical spectrum
indicating a Type 1 (broad-line) AGN, and therefore need not be
classified using this narrow-line diagnostic.  IRAS~23365+3604 fails
the signal-to-noise cut for the \niilam\ line but occupies the AGN
portion of the \oi~$\lambda$6300/\ha\ diagram.  Finally, Arp~299A \&
Arp~299B are both classified as SF/AGN based on the classification
of the Arp~299 system.  With the exception of IRAS~12112+0305 and
VII~Zw~31, spectra were obtained with a long-slit drift-scanning
technique in order to integrate the light spatially (MK06).
Spectroscopic measurements for IRAS~12112+0305 and VII~Zw~31 were
obtained from \citet{1999ApJ...522..113V} and from
\citet{1998A&AS..132..181W}, respectively.

We note that dust obscuration can challenge the identification of 
an AGN in the optical spectral range.  Sufficient optically thick 
material with a large covering fraction surrounding the nuclear 
region could potentially mask all AGN signatures at these wavelengths.  
However, we identify an AGN in all but one of the most IR-bright
galaxies (IRAS~17208-0014).  The latter is the only optically-obscured, 
potentially MIR-obscured, AGN candidate sometimes referred to as 
{\it buried} AGN \citep{2009ApJ...694..751I}.  
We also note that although this diagnostic indicates the presence of an
AGN, it does not allow to quantify its strength and its relative
contribution to the total infrared light of its host galaxy.  For
example, the well-known ULIRG Arp~220 is thought to be mainly powered
by star formation \citep{1996A&A...315L.137L} even though it is known 
to host a Seyfert nucleus \citep{1985ApJ...290..116R}.
Furthermore, there has been accumulating evidence that most LIRGs 
and ULIRGs are mainly powered by starbursts 
\citep{1998ApJ...505L.103L, 1998ApJ...498..579G, 2003MNRAS.343..585F, 
2008A&A...484..631V, 2008MNRAS.385L.130N}, except for a few notable 
cases such as Mrk~231 \citep{1991ApJ...378...65C, 2007ApJ...654L..49S}, 
IRAS~05189-2524 \citep{2007ApJ...654L..49S, 2007ApJ...656..148A}, and
NGC~1068 \citep{2001A&A...367..487L}.  These three examples stand out 
from the rest of our sample as they exhibit the largest mid-IR excesses 
(they correspond to the three blue triangles below the dotted line 
in Figure~\ref{fig:FIR_IR}).  

Among others, \citep{2008MNRAS.385L.130N} report that 
the AGN contribution may dominate in the mid-IR range while remaining 
a small fraction of the total infrared light, typically less than 25\%.
Recent work suggests a correlation between the optical spectral 
classification and the AGN contribution to the bolometric luminosity of
ULIRGs.  Using a sample of 74 ULIRGs, \citet{2009ApJS..182..628V} find 
that the AGN contribution ranges from $\sim$ 15\%$-$35\% among purely star-forming 
(SF) and LINER ULIRGs to $\sim$50 and 75\% among Seyfert 2 and Seyfert 1 ULIRGs, 
respectively.  Individual galaxies may deviate from this trend as their sample
includes a Seyfert 2 ULIRG with as little as 20\% AGN contribution and SF-classified 
ULIRGs with roughly 50\% AGN contribution.  Thus, we stress that although the 
optical classification used here indicates the presence of an AGN, it does not 
allow us to quantify its contribution at infrared wavelenghts on a 
galaxy-by-galaxy basis.  Indeed, several systems classified as AGN are also 
undergoing major episodes of star formation.

\subsection{Infrared Luminosity}
\label{sec:FIR_IR}

We estimate far-infrared [$L_{FIR}\equiv L(40-500~\mu$m$)$] and total
infrared [$L_{IR}\equiv L(8-1100~\mu$m$)$] luminosities of the galaxies 
in our sample using IRAS observations at 12, 25, 60, and 100~$\mu$m
\citep{2003AJ....126.1607S}.  Following \citet{2006ApJ...642..775M}, 
we model the infrared spectral energy distribution of each
object longward of 100 micron using a modified blackbody with dust
emissivity proportional to $\lambda^{-1}$ and dust temperature given by 
the observed $S_{\nu}(60~\micron) / S_{\nu}(100~\micron)$ flux ratio 
\citep{2000ApJ...533..236G, 2003ApJ...586..794B}.  We then integrate 
numerically over the appropriate wavelength range to derive 
$L_{FIR}$ and $L_{IR}$.

\begin{figure}
\includegraphics[clip=true,width=\linewidth]{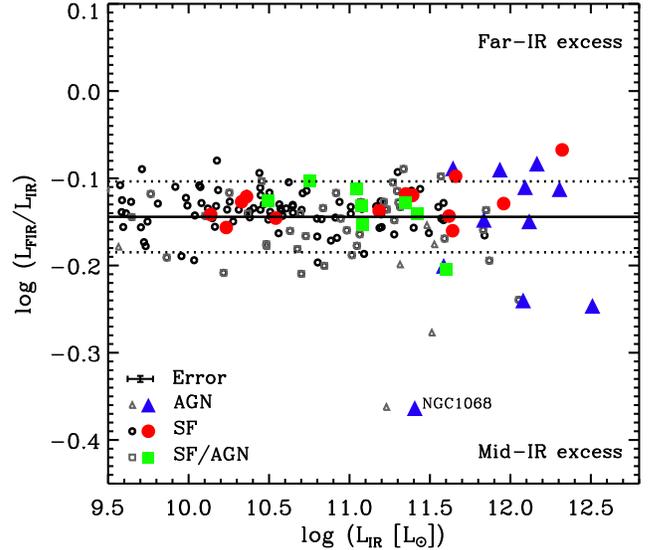}
\caption{FIR/IR luminosity ratio for the MK06 sample of galaxies (gray
  and black open symbols) and for our combined sample of galaxies (filled
  colored symbols as in Figure~\ref{fig:BPT}, also see legend).  We
  find an average value of log($L_{FIR}/L_{IR}$) = -0.14
  (corresponding to $L_{IR}/L_{FIR} = 1.38$, which is close to the
  value of 1.3 used in GC08).  Although most of the galaxies in our
  sample are fairly close to that value, there are some noticeable
  outliers, the most striking one being NGC~1068. We stress that our
  results remain unchanged (within the uncertainties) if we adopt FIR
  instead of total IR luminosities in our analysis.
  (A color version of this figure is available in the online journal.)}
\label{fig:FIR_IR}
\end{figure}

When interpreting galaxy infrared luminosity, one has to take into
account potential contributions of recent or ongoing star formation,
older stellar populations, and AGN.  In the case of LIRGs and ULIRGs,
the infrared emission from dust heated by powerful starbursts
dominates over that contributed by old stars, so we ignore the older
stellar population component.  It has been suggested that because 
the typical AGN IR SED peaks in the mid-infrared range 
\citep[around $\lambda = 10-20~\mu$m, e.g.][]{1994ApJS...95....1E}, 
the far-infrared portion of the SED roughly corresponds to the
star-forming component.  We compare the total IR and FIR
emission in Figure~\ref{fig:FIR_IR} and find a similar ratio
($L_{IR}/L_{FIR} = 1.38$) to that reported in the literature (1.3,
GC08).  Most of the galaxies in our sample lie within one sigma of the
mean value obtained for a control sample of galaxies (MK06, small 
plotting symbols in Figure~\ref{fig:FIR_IR}).  There
are a few outliers, the most striking example being the prototype
Seyfert 2 galaxy NGC~1068, which has the largest MIR-excess and lies
4.5$\sigma$ from the mean.  Interestingly, even when the fractional AGN 
contribution to the mid-IR light dominates over the starburst contribution, 
it still appears to be the case that the AGN total IR contribution is 
mostly less than 25-30\% \citep{2008MNRAS.385L.130N}.

The larger spread in $L_{IR}/L_{FIR}$ ratios observed in brighter 
galaxies may be related to a combination of varying extinction and 
AGN contribution.  
For instance, \citet{2009ApJS..182..628V} suggest an evolutionary 
sequence with three AGN classes for nearby ULIRGs.  The first class is 
characterized by small extinctions and large PAH equivalent widths.  
ULIRGs in this class are highly starburst-dominated. 
The second class includes galaxies with large extinctions but that 
are still likely dominated by starburst; while galaxies belonging in 
the third class show both small extinctions and PAH equivalent widths 
and have a significant AGN contribution (at least as important as the 
starburst).  A detailed analysis is beyond the scope of this paper,
but we refer the reader to published analyses utilizing mid-infrared 
spectroscopy \citep[e.g.][]{1998ApJ...505L.103L, 1998ApJ...498..579G, 
2007ApJ...654L..49S, 2007ApJ...656..148A, 2007ApJ...667..149F}.
Here, we simply identify the presence of an AGN using optical 
emission lines (\S\ref{sec:BPT}) and we do not attempt to quantify 
its contribution, given the large uncertainties in doing so.

Because we do not interpret $L_{IR}$ strictly as a SFR, we choose to
plot the total infrared luminosity throughout this paper.  We repeated
our analysis using FIR instead of IR luminosity.  This change has a
negligible effect on our results (the differences are all well within the
uncertainties, see Tables~\ref{tab:slopes} \& \ref{tab:ratios}).

\subsection{Numerical Models}
\label{sec:meth_model}
We utilize the numerical models of \citet{2008ApJ...684..996N} as a
comparison for our observations. We refer the reader to this
work for details on the models, though we summarize the aspects most
relevant to the present study.

In an effort to model the observed SFR-\lmol \ relations in galaxies,
\citet{2008ApJ...684..996N} coupled 3D non-LTE radiative transfer
calculations with smoothed particle hydrodynamics (SPH) simulations of
galaxies in evolution. The SPH simulations were calculated utilizing a
modified version of the publicly available code GADGET-2
\citep{2005MNRAS.364.1105S}, including prescriptions for a
multi-phase ISM, supernovae pressurization of the ISM, and
star-formation following a generalized (3D) version of the
Kennicutt-Schmidt law \citep{2003MNRAS.339..289S}. Here, we set the
index of the Kennicutt-Schmidt relation to 1.5, which has
important consequences in driving the simulated SFR-\lmol \ relations
\citep{2008ApJ...684..996N}. We additionally include energy feedback
from accreting AGN \citep{2005MNRAS.361..776S}, though note that it
has negligible impact on the simulated SFR-\lmol \ relations.

The molecular line emission properties of the model galaxies were
extracted using the 3D non-LTE radiative transfer code, {\it
  Turtlebeach} \citep{2006ApJ...642L.107N}. {\it Turtlebeach}
considers both collisional and radiative (de-)excitation in determining
the level populations of a given molecule, and utilizes Monte Carlo
methods for sampling the spatial and frequency domains. Because the
hydrodynamic simulations typically have a coarser physical resolution 
than the scale of GMCs, sub-grid prescriptions for including GMCs as
singular isothermal spheres following a Galactic mass spectrum and
mass-radius relation have been implemented \citep[for details, please
  see][]{2006ApJ...642L.107N, 2008ApJS..176..331N}.

The SPH simulations consist of $\sim$100 galaxies comprised of
isolated disks as well as gas-rich, binary, 1:1 galaxy mergers. The
structure of the galaxies were initialized following the
\citet{1998MNRAS.295..319M} formalism. In order to probe a relatively
large dynamic range of galaxies, the galaxies were initialized with
gas fractions $f_{\rm g}$[0.2,0.4,0.8] and total (halo) mass ranging
from $\sim$1$\times$10$^{12} M_\odot$ to $\sim$4$\times$10$^{13}
M_\odot$ spaced in 4 mass bins. The galaxy mergers were run at a
single initial gas fraction and mass ($f_{\rm g}$=0.4 and $M_{\rm
  DM}$=1$\times$10$^{12} M_\odot$). A key feature of these simulations
is that they include constant Galactic abundances for HCN and CO, i.e. no
chemistry is modeled. These simulations were shown to 
accurately recover the observed $L_{\rm IR}$-\hcnOne, $L_{\rm
  IR}$-\coOne \ and $L_{\rm IR}$-\coThr \ relations.  Furthermore, 
they predicted a sub-linear $L_{\rm IR}$-\hcnThr \ relation,
which was subsequently observed by B08.  

The HCN and CO simulations were taken from the study of
\citep{2008ApJ...684..996N}, while HCO$^+$ simulations of 
gas-rich galaxy mergers were run specifically for this work. 

\section{Results}
\label{sec:results}

\subsection{Correlation Between Molecular Line and Infrared Luminosities}
\label{sec:lum}

\begin{figure}
\includegraphics[clip=true,width=\linewidth]{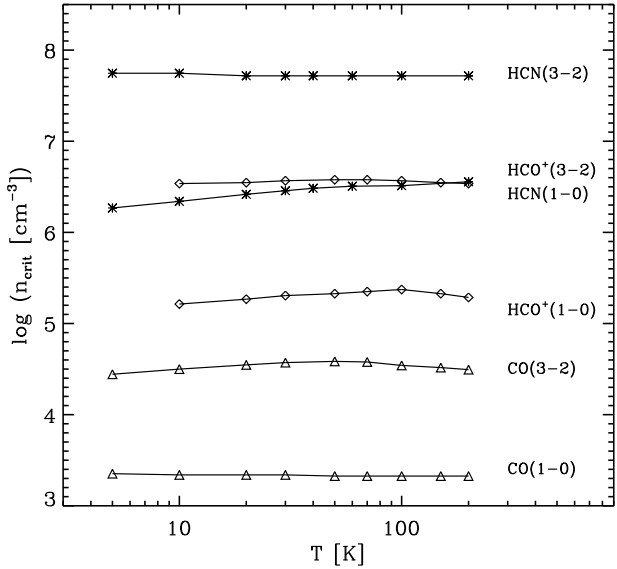}
\caption{Critical density, $n_{crit} \equiv A_{ul}/\Gamma_{ul}$, for
  typical molecular transitions used to trace molecular gas in
  galaxies.  The critical density is fairly insensitive to temperature
  over a wide range.  Note that the \hcnOne\ and \hcoThr\ transitions
  have a remarkably close value of critical density, whereas the critical
  density of \hcoOne\ is over one order of magnitude smaller than that of
   \hcnOne.  Molecular data come from the Leiden Atomic and Molecular 
  Database \citep{2005A&A...432..369S}.}
\label{fig:ncrit}
\end{figure}

The molecular gas tracers used in this work span a broad range of
critical densities 
($n_{crit} \equiv A_{ul}/\Gamma_{ul}$) varying between $\sim 10^{3.3}
~{\rm cm}^{-3}$ for \coOne\ and $\sim 10^{7.7}~{\rm cm}^{-3}$ for
\hcnThr\ (Figure~\ref{fig:ncrit}).  Although the critical density of
\hcnOne\ is an order of magnitude larger than that of \hcoOne\, it is
almost a perfect match to the critical density of \hcoThr. In what
follows, we consider the values of $n_{crit}$ at an assumed
temperature of 30~K, noting that they are roughly constant over a 
broad range of temperatures ($20 - 100$~K, see Figure~\ref{fig:ncrit}).
Even though we use $n_{crit}$ to guide some of our interpretations, we
caution that molecular gas emission can be observed in gas with densities
less than critical \citep{1999ARA&A..37..311E}.  Substituting critical
densities with effective densities as defined in \citet{1999ARA&A..37..311E}
does not alter our main conclusions.

\begin{figure*}
\epsscale{0.75}
\plotone{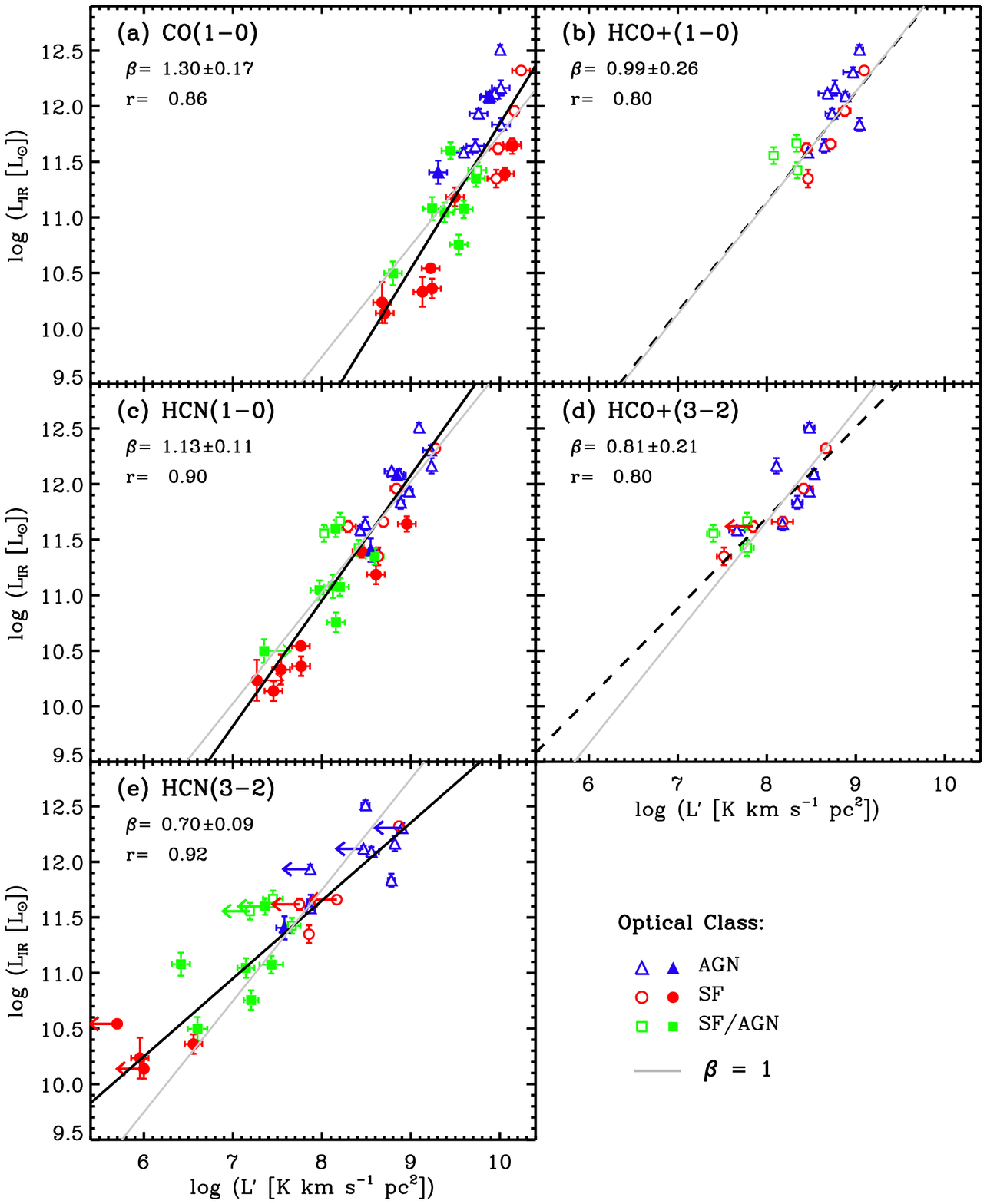}
\caption{Relationship between infrared luminosity and  \coOne\ {\it(a)}, 
  \hcoOne\ {\it(b)}, \hcnOne\ {\it(c)}, \hcoThr\ {\it(d)},  and 
  \hcnThr\ {\it(e)},  molecular line luminosities.  We show
  galaxies from our primary sample (filled symbols) as well as from
  the GC08 sample (open symbols).  Panels are ordered by increasing
  molecular line critical density.  Our optical AGN classification is
  shown using the same color scheme as in Figure~1 (also see legend).
  Each panel is labeled with the mean and standard deviation of the 
  corresponding slope ($\beta$) distribution and with the linear 
  correlation coefficient $r$. The gray lines have a slope of unity
  and are included for visualization purposes.
  (A color version of this figure is available in the online journal.)}
\label{fig:linelum}
\end{figure*}

Assuming that $L_{\rm IR}$ (or $L_{\rm FIR}$) provides a good estimate of SFR 
and that the molecular line luminosity traces the mass of molecular gas above a 
certain density, $L_{\rm IR}-L'_{mol}$ relations (or their surface density equivalent) 
are commonly interpreted and/or used to derive universal SFR prescriptions.  
In this section, we present relationships between galaxy infrared
luminosity and the luminosity of various molecular lines.  We compute
the slope ($\beta$) between log($L_{IR}$) and molecular line
luminosity of \coOne, \hcoOne, \hcoThr, \hcnOne\ and
\hcnThr\ (Figure~\ref{fig:linelum}).  When available, we compute the
slope for the combined sample (solid lines).  However, observations for
the \hcoOne\ and \hcoThr\ transitions are only available for the GC08
sample (open symbols).  The latter sample comprises fewer galaxies 
and is biased towards higher infrared luminosities, resulting in derived 
slopes (dashed lines) with much larger uncertainties.
Galaxies hosting an AGN (blue triangles) occupy the bright end of 
$L_{IR}$ but appear to follow the same relationship as the rest of 
the sample.  Molecular line non-detections are shown with arrows but 
are not included when computing the slope ($\beta$) or the Pearson 
correlation coefficient (r).

All slopes are computed using the LINMIX IDL routines from
\citet{2007ApJ...665.1489K}. In these routines, the distributions 
of independent variables are modeled as a mixture of Gaussian 
functions, which allows for greater flexibility when computing the true 
distributions of these variables (i.e. without measurement errors)
given the observations.  Using a Bayesian statistical approach, 
the likelihood distribution function is computed 
and then integrated over the entire data set.  This method offers
several advantages compared to other algorithms. Namely, it provides
likelihood distributions for the values of slopes and intercepts while
allowing for intrinsic scatter (i.e. scatter present in the absence
of measurement errors).  Most relevant for this work, this method is
successful at recovering linear regressions when measurement errors
dominate the scatter and when there is a non-negligible number of
non-detections in the sample.  We note that the value of the slope found
using a simple linear least-squares fitting routine available in IDL 
(LINFIT) is always included within one-sigma of the mean of the distribution 
of slopes derived by LINMIX.  The mean values of the distribution of slopes
are shown in Figure~\ref{fig:linelum}.  We also report the slopes obtained 
by substituting $L_{IR}$ by $L_{FIR}$ (Table~2).  The FIR slopes are
slightly higher on average, but consistent within $1\sigma$.
Our results agree with previous work in that $\beta_{CO(1-0)} > 1$ 
(2$\sigma$) and that $\beta_{HCN(3-2)} < 1$ (3$\sigma$).
The derived slopes for \hcoOne, \hcnOne\ and \hcoThr are statistically 
consistent with each other (and with a slope of unity), given the large 
uncertainties.

\begin{figure*}
\centering
\includegraphics[clip=true,width=170mm]{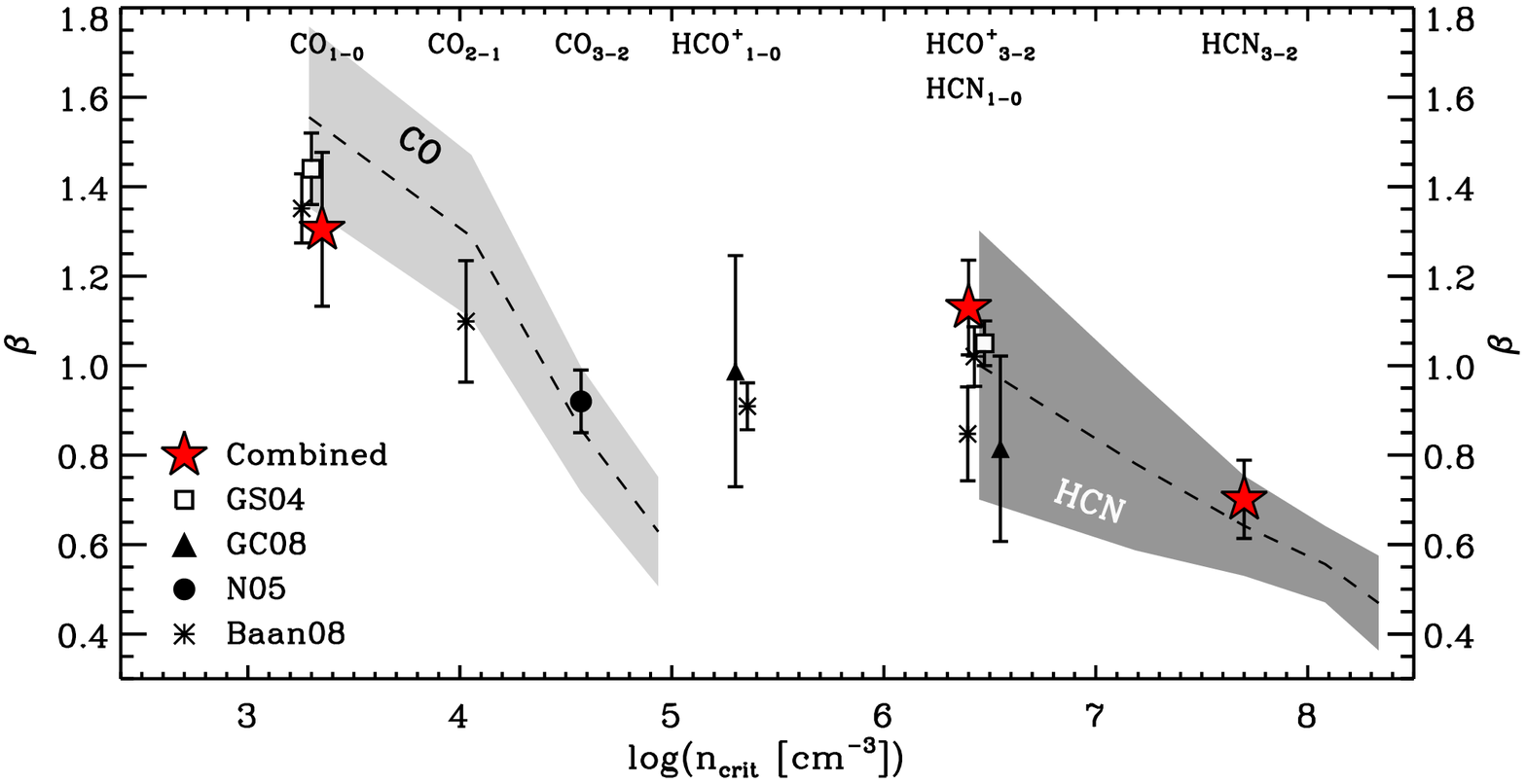}
\caption{ Slope of log($L_{IR}$) $-$ log($L'$) versus molecular line
  critical densities for the GC08 sample (triangles) and our combined
  sample (stars).  Individual slopes are shown in the previous figure.
  We add values from the literature: \coOne\ and \hcnOne\ from GS04
  (open squares), CO($3-2$) from \citet[filled
    circle]{2005ApJ...630..269N} and, in order of increasing critical
  density, \coOne, CO($2-1$), \hcoOne, CS($3-2$) and \hcnOne\ from
  \citet[asterisk symbols]{2008A&A...477..747B}.  The corresponding
  molecular line transitions are labeled at the top of the figure.
  Some of the points around log($n_{crit}$) $\sim$ 3.3 and 6.5 were
  offset slightly in critical density for clarity of the plotting
  symbols.  The shaded regions show the predictions of N08 models for
  CO (pale gray) and HCN (dark gray) rotational transitions from
  $J=1-0$ to $J=5-4$.  See text for more detail. 
  (A color version of this figure is available in the online journal.)}
\label{fig:slopes}
\end{figure*}

Figure~\ref{fig:linelum} suggests that the proportionality between
infrared luminosity log($L_{IR}$) and molecular line luminosity
log($L'$) varies as a function of critical density.  We compile the
mean and standard deviation of slope distributions ($\beta \pm
\sigma$) obtained previously and show them as a function of the 
critical density of the corresponding molecular transition 
(Figure~\ref{fig:slopes}).  We find evidence
for a shallower log($L_{IR}$) $-$ log($L'$) slope with increasing
molecular line critical density. We supplement our
results with published relations between log($L_{IR}$) and
log($L'_{mol}$) for the following transitions: \coOne\ and
\hcnOne\ from GS04 (open squares), CO($3-2$) from \citet[filled
  circle]{2005ApJ...630..269N} and, in order of increasing critical
density, \coOne, CO($2-1$), \hcoOne, CS($3-2$) and \hcnOne\ from
\citet[asterisk symbols]{2008A&A...477..747B}.
\citet{2008A&A...477..747B} published slopes for the inverse relation
[log($L'_{mol}$) $-$ log($L_{IR}$)].  Thus, the asterisks symbols are
not truly calculated slopes, but rather the inverse of the published
numbers. These points are included for visual reference, but should be
considered more uncertain than the associated error bars.  All
transitions are labeled at the top of the figure.  Points
corresponding to \hcnOne\ and HCO$^+$(3-2) are very close given
their near-identical critical densities.  We added small offsets 
in critical density for clarity of the plotting symbols around
log($n_{crit}$) $\sim$ 3.3 and 6.5. 

Similar results have been found by \citet{2009ApJ...695.1537I}.  
These authors looked at the reciprocal relation but find a slope 
that varies with critical density in the same sense (i.e. the inverse 
of their slope decreases with $n_{crit}$, reaching close to unity 
for their higher density tracers). While they only considered \coOne,
CO($3-2$) and HCN($1-0$), here we extend the trend to a larger
number of molecular line transitions and, more importantly, to
a transition with a higher critical density probing the regime
where the log($L_{IR}$) $-$ log($L'$) slope is less than unity
[\hcnThr].

In their coupling of non-LTE radiative transfer calculations
with hydrodynamic galaxy evolution simulations, 
\citet{2008ApJ...684..996N} predicted the SFR-\lmol \ relations 
in local galaxies as a function of increasing
molecular transition which serve as a natural comparison for the
observations compiled here. We compare the observations with the
numerical simulations of Narayanan et al. in Figure~\ref{fig:slopes}
by over-plotting their model predictions as the gray shaded
regions\footnote{We note that \citet{2008ApJ...684..996N} only
  published the model SFR-\lmol \ slopes for CO and HCN.}. 
The predicted slopes were calculated by randomly sampling 25 of 
the $\sim$100 model galaxies in the Narayanan et al. sample, and 
calculating the slope for the subset of galaxies. The shaded region 
is the 1$\sigma$ dispersion in slopes, and the dashed line is the 
mean. The model SFR-\lmol \ slope decreases as a function of 
increasing critical density owing to the arguments described in 
\S 1. Because higher J-level molecular line transitions trace a 
smaller fraction of the galaxy's molecular content, the SFR-\lmol\ 
slope decreases from the assumed Kennicutt-Schmidt index of 1.5.  We
find good agreement between the compiled molecular line observations
and the theoretical predictions for CO and HCN.

\subsection{Comparison Between High- and Low-Density Molecular Tracers}
\label{sec:lineratios}

In this section, we study the possibility that the enhanced
\hcnOne/\coOne\ luminosity ratio actually corresponds to an enhanced
dense gas fraction.  The five transitions studied in this work allow
us to examine ten molecular line luminosity ratios as a function of IR
luminosity.  This set includes published line ratios for which the
lines involved probe different densities.  We introduce a new line
ratio, \LphcoThr/\LphcnOne, for which both molecular transitions have 
nearly the same critical density (see Figure~\ref{fig:ncrit}).

When comparing two tracers, we adopt the convention of dividing 
the higher-density (HD) tracer by the lower-density (LD) tracer.
We quantify the contrast of their critical densities as follows:
$R_{crit} \equiv$ log$(n_{crit}^{HD}/n_{crit}^{LD})$.  Our set of line
ratios spans 4 orders of magnitude in critical density contrast $0.1 <
R_{crit} < 4.4$.  We quantify the IR luminosity dependence of each
molecular line ratio by a power-law (see Figure~\ref{fig:ratio_lir}).  
Whenever possible, we fit for our combined sample (open and filled
symbols, solid lines).  Otherwise, we fit to the GC08 sample only
(open symbols, dashed lines).  We report the mean and standard
deviation on the distribution of possible slopes ($\alpha$) in the top
of each panel.  We also include the critical density contrast
($R_{crit}$) and the Pearson linear correlation coefficient ($r$) in
each case.

\begin{figure*}
\epsscale{0.85}
\includegraphics[clip=true,width=\linewidth]{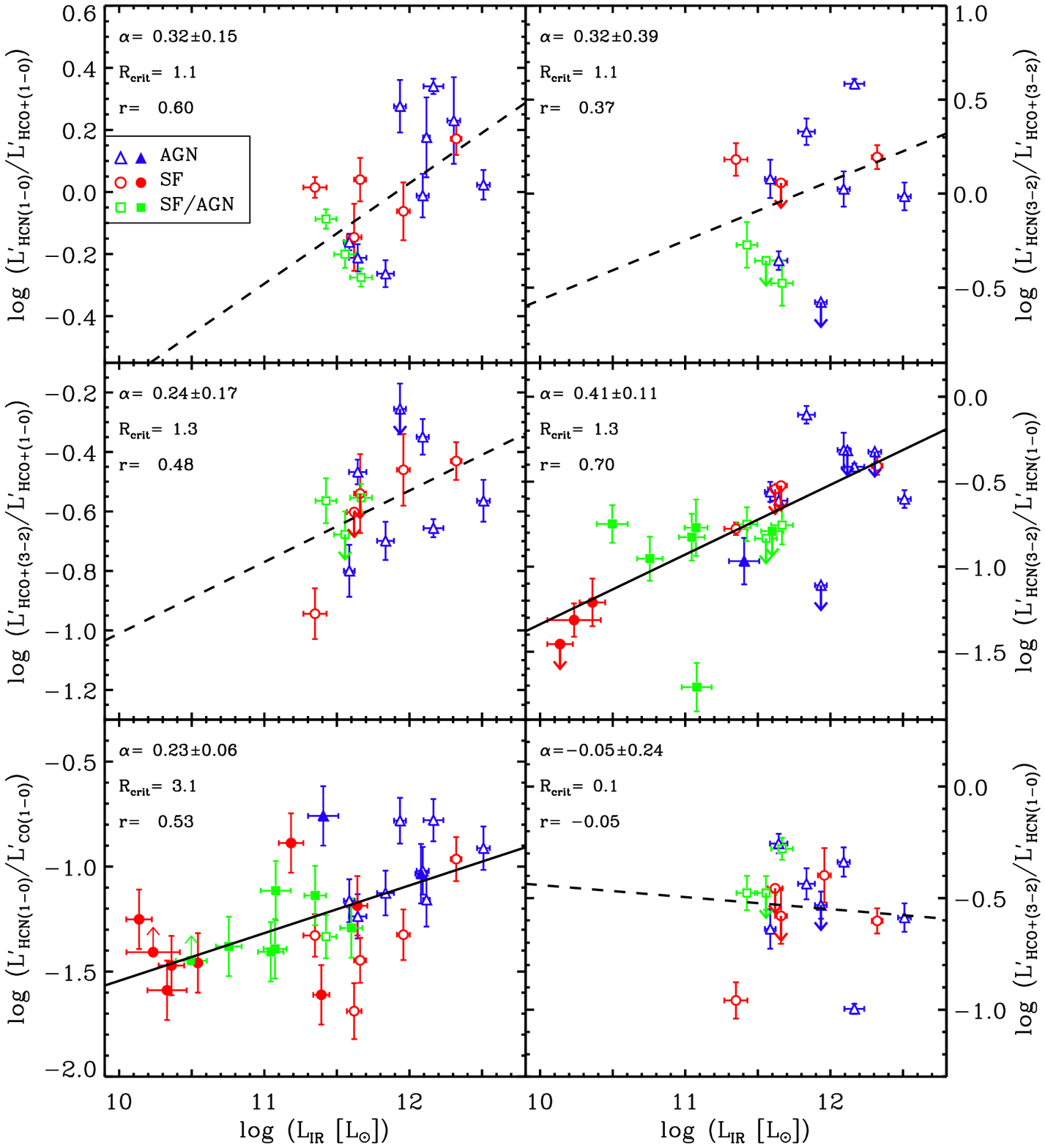}
\caption{Molecular line luminosity ratios as a function of infrared
  luminosity. We compare high- and low-critical density molecular
  lines, log($L'_{HD}/L'_{LD}$), and label each panel with the
  best-fit slope ($\alpha$), the ratio of the critical densities
  ($R_{crit} \equiv$ log$(n_{crit}^{HD}/n_{crit}^{LD})$) and the linear
  correlation coefficient ($r$).  The slope
  is computed for our combined sample when available (solid lines) and
  otherwise for the GC08 sample only (dashed lines).  The line ratios
  used are: {\it Top left}: HCN$_{J=1-0}$/HCO$^{+}_{J=1-0}$, {\it Top
    right}: HCN$_{J=3-2}$/HCO$^{+}_{J=3-2}$, {\it Middle left}:
  HCO$^{+}_{J=3-2}$/HCO$^{+}_{J=1-0}$, {\it Middle right}:
  HCN$_{J=3-2}$/HCN$_{J=1-0}$, {\it Bottom left}:
  HCN$_{J=1-0}$/CO$_{J=1-0}$, {\it Bottom right}:
  HCO$^{+}_{J=3-2}$/HCN$_{J=1-0}$.
  The different symbols represent the optical spectral type: star-forming
  (SF) ({\em red circles}); AGN ({\em blue triangles}); and SF/AGN
  ({\em green squares}).  Open symbols denote galaxies from the GC08
  sample, whereas filled symbols are used for galaxies in our primary 
  sample only.
  (A color version of this figure is available in the online journal.)}
\label{fig:ratio_lir}
\end{figure*}

\begin{figure*}
\centering
\addtocounter{figure}{-1}
\includegraphics[clip=true,width=\linewidth]{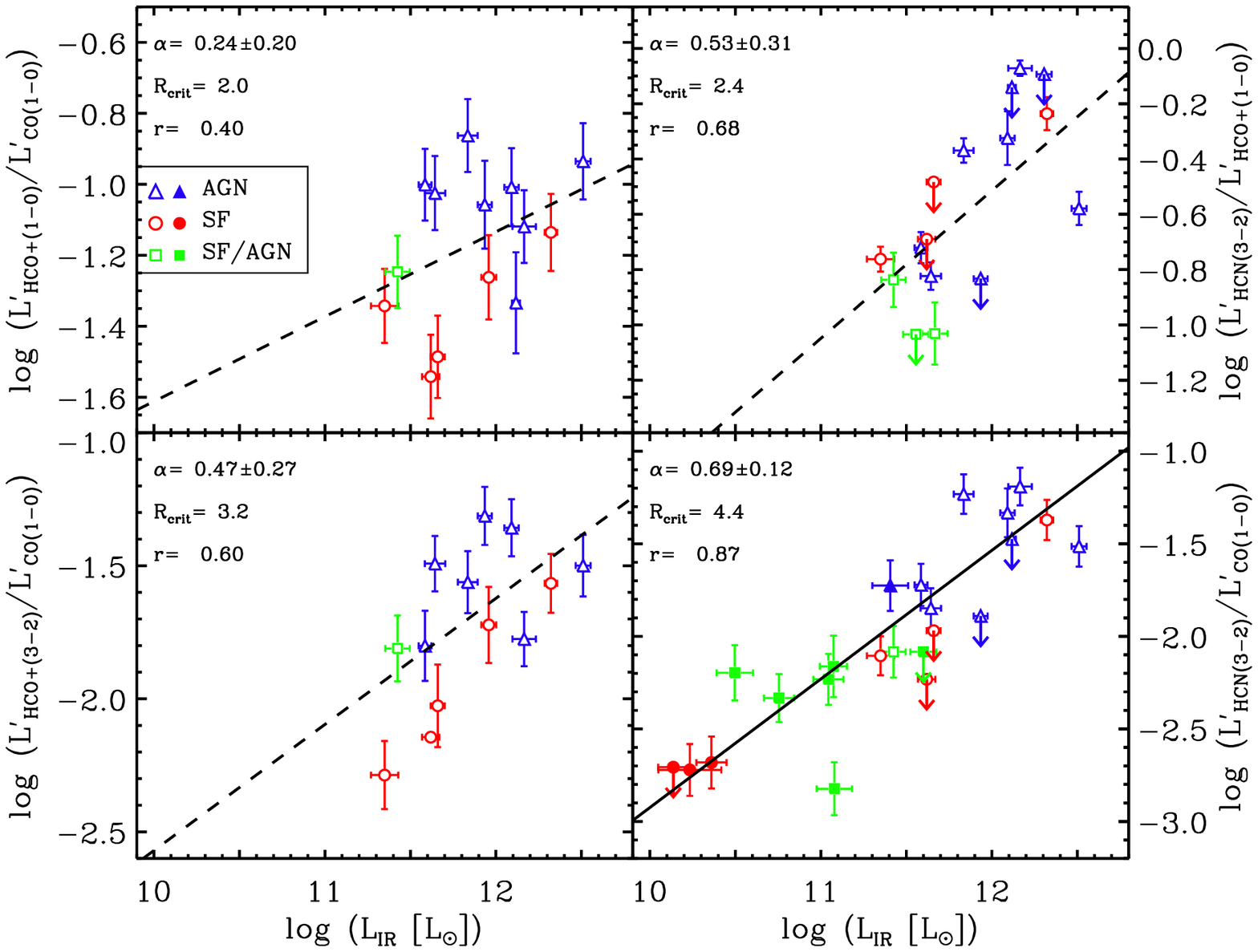}
\caption{continued.  Ratios shown here are:
  {\it Top left}: HCO$^{+}_{J=1-0}$/CO$_{J=1-0}$, 
  {\it Top right}: HCN$_{J=3-2}$/HCO$^{+}_{J=1-0}$, 
  {\it Bottom left}: HCO$^{+}_{J=3-2}$/CO$_{J=1-0}$,
  {\it Bottom right}: HCN$_{J=3-2}$/CO$_{J=1-0}$.
  (A color version of this figure is available in the online journal.)}
\end{figure*}

Our results are summarized in Table~\ref{tab:ratios} and
indicate an enhancement in molecular luminosity ratio of high- to
low-density tracer ($L'_{HD}/L'_{LD}$) with increasing IR luminosity.
This trend is observed, with varying degree of significance, for all
ratios with $R_{crit} > 1$. In contrast, \LphcoThr/\LphcnOne\ --- the only
ratio of lines with nearly equal critical densities ($R_{crit} \sim
0$) --- remains flat even at high infrared luminosities.  The same trend
is observed when substituting $L_{IR}$ by $L_{FIR}$ (see
Table~\ref{tab:ratios}).

The relations are better constrained ($> 3\sigma$) for transitions for
which it is possible to use our combined sample as it increases the
dynamic range in IR luminosity by one order of magnitude.  Cases where 
only the GC08 observations are available are subject to more uncertainty, 
and would benefit from additional measurements at lower IR luminosities 
to confirm the trend observed here.  Nevertheless, we find a compelling
case for interpreting the variations in line luminosity ratios in
terms of the molecular gas density distribution.  

\begin{figure}
\includegraphics[clip=true,width=\linewidth]{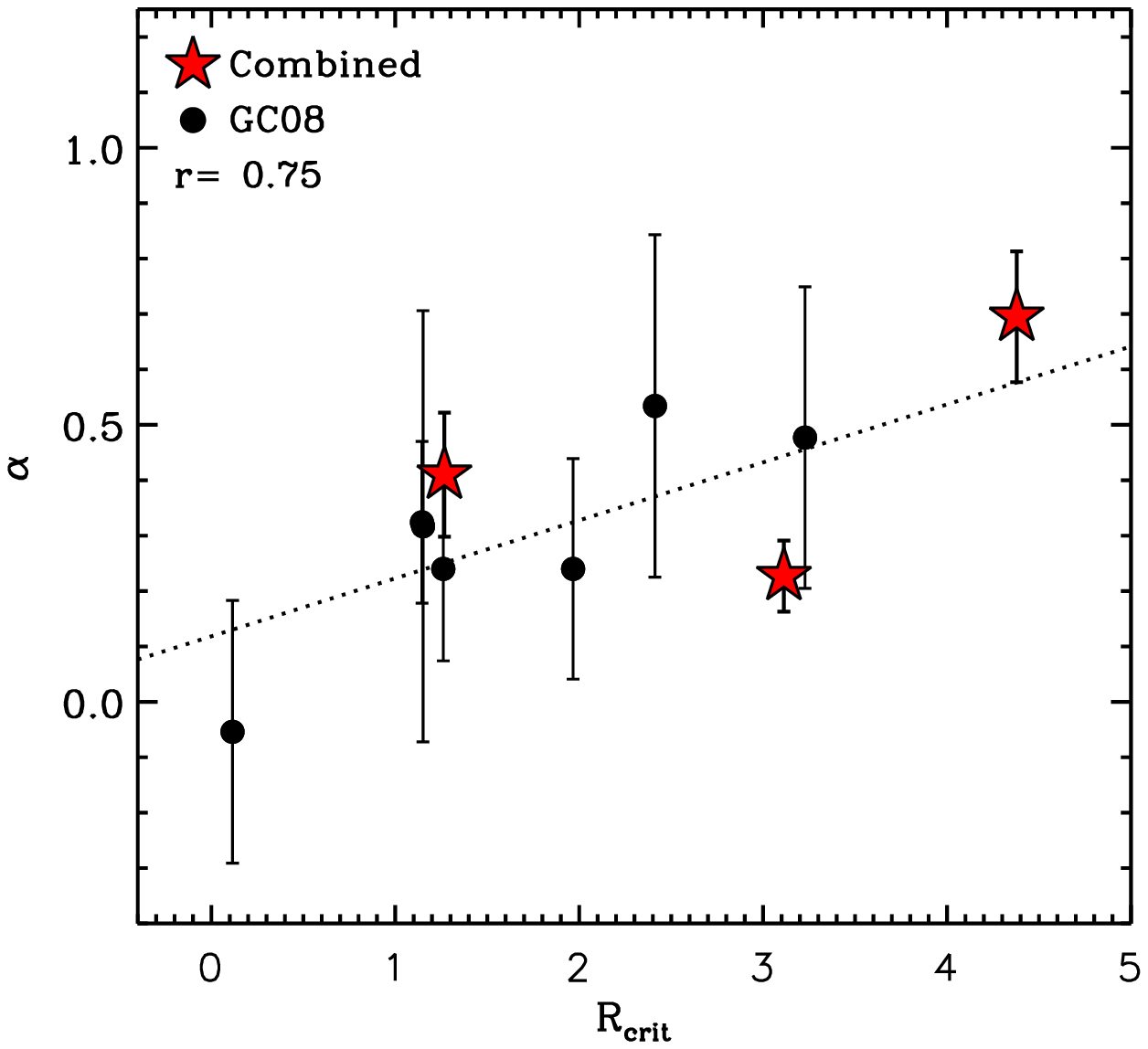}
\caption{Values of $\alpha$ as a function of $R_{crit}$, the logarithm
  of the ratio of critical densities of the lines.  The index $\alpha$
  characterizes the $L_{IR}$-dependence of the molecular line
  luminosity ratios shown in Figure~\ref{fig:ratio_lir}
  (i.e. ($L'_{HD}/L'_{LD}$) $\propto$ ($L_{IR}$)$^{\alpha}$).  Most
  line ratios included in this figure have $R_{crit} > 1$.
  Correspondingly, these ratios show a positive index indicating an
  increase toward high $L_{IR}$.  On the other hand, \hcnOne\ and
  \hcoThr\ have a similar critical density (their $R_{crit}$ is close to
  zero), and their luminosity ratio is consistent with being constant
  with infrared luminosity ($\alpha$ = 0).  The slope distribution of
  the points in this figure has a mean and standard deviation of 
  $0.10 \pm 0.07$ ($1.5\sigma$ away from zero, dotted line).
  (A color version of this figure is available in the online journal.)}
\label{fig:ratios_slopes}
\end{figure}

In Figure~\ref{fig:ratios_slopes}, we present a compilation of the values
of $\alpha$ shown in Figure~\ref{fig:ratio_lir}.  We identify the
points that originate from our combined sample (red stars) and those
that correspond to the GC08 sample (black circles).  Although we
expect other variables such as molecular gas temperature and
excitation to add to the scatter around each value of $\alpha$, the
existence of a significant positive correlation in
Figure~\ref{fig:ratios_slopes} suggests that the molecular gas density
distribution may be the primary physical mechanism driving the observed 
molecular line luminosity ratios in LIRGs and ULIRGs.

This result is consistent with the analysis of \citet{2009ApJ...695.1537I} 
who find that the CO($3-2$) source size is more compact than the CO($1-0$) size for ULIRGs 
suggesting that their high FIR (IR) luminosity is linked with them 
having a large amount of dense molecular gas concentrated within 
their central region.  Furthermore, there is evidence suggesting 
that the ISM ambient density in ULIRGs is higher by a factor of 100 
compared to normal star-forming galaxies \citep{1997ApJ...478..144S}.
A number of other studies support the presence of warm and dense molecular 
gas in these IR-bright galaxies \citep[e.g.][]{2007ApJ...659..296L, 
2007ApJ...656..148A}.

Our findings are also in agreement with the Arp~220 and NGC~6240 case 
studies of \citet{2009ApJ...692.1432G}, who report an increased dense 
gas fraction in these two prototypical ULIRGs. Using measurements of 
a large number of dense molecular tracers, these authors infer that, 
for these two ULIRGs, most of the molecular gas is in a dense phase 
and that the GMCs mass-size power-law is steeper than in normal star-forming 
galaxies, indicating that HCN emission traces denser and more compact GMCs 
compared to HCO+ transitions between the same J levels.

\subsection{Possible Causes of Higher Gas Density}

Assuming that variations in temperature and excitation are smaller than the 
effect of varying density, a positive relation between $\alpha$ and $R_{crit}$ 
would imply that more IR-luminous galaxies have a larger fraction of 
dense gas than galaxies with smaller IR luminosities (see \S\ref{sec:model} 
for comparison to radiative transfer models). Several mechanisms 
could create such a situation (which translates to a larger average density 
of the molecular gas). For example, superwinds capable of expelling 
interstellar material have been observed in IR-bright starburst galaxies 
\citep{2000ApJS..129..493H}. AGN-driven outflows have been modeled and are
predicted to be capable of expelling molecular gas from galaxies 
\citep{2006ApJ...642L.107N, 2008ApJS..176..331N}.  

Although signatures of atomic gas outflows have been reported in a
large number of studies \citep[e.g.][for a review, see \citet{2005ARA&A..43..769V}]
{1990ApJS...74..833H, 1995ApJS...98..171V, 2005ApJS..160..115R, 
2006ApJ...647..222M, 2009arXiv0907.4370S}, observational evidence for 
large-scale outflows of molecular gas is scarce and very recent.  Based on 
their observations of CO(3-2), HCO$^{+}$(3-2), HCO$^{+}$(4-3), 
\citet{2009ApJ...700L.104S} find P-cygni profiles indicating outflows of 
molecular gas around the two merging nuclei of Arp~220.  They also report 
that the HCO$^+$ emission is more concentrated around the nuclei relative to 
the CO emission, in agreement with the picture presented here, where this 
difference would reflect an enhanced molecular gas density in these regions.
In addition, molecular gas outflows have recently been identified for the S0 galaxy 
NGC~1266 (Alatalo et al., in prep).  In this case, the outflow may be associated
with the high-density enshrouded nucleus \citep{1998A&A...334..482M}, but the 
powering source (AGN or stellar) remains unclear. Future searches for molecular 
outflows in galaxies would be useful to understand their energy and gas 
content balance.

The gas density distribution could be affected if, for example, low-density 
gas is depleted more easily than denser gas during such outflows.  In support 
of this scenario, lower-density medium is expected to occupy a larger volume 
and would acquire larger outflow velocity by conservation of momentum.
In addition, the evidence that outflow velocities increase with IR luminosity
\citep{2005ApJ...621..227M, 2005ApJS..160..115R} is consistent, at least
qualitatively, with an increase in dense gas fraction in more
IR-luminous galaxies as they can more easily deplete less dense gas.
On the other hand, gas accretion and inflows could counteract this 
effect to a certain extent and help replenish the gas content.

Alternatively, one could posit that galaxies with a larger reservoir
of dense gas (and/or higher dense gas fraction) can sustain higher IR
luminosity by allowing for more powerful starbursts and AGN activity.
Mergers of gas-rich galaxies can trigger and fuel both starburst and
AGN activity and consequently appear as ULIRGs.  Indeed, most if not all
of the ULIRGs in this sample are undergoing a galaxy merger.
The important change in the dynamics of these systems, and in
particular, of their molecular gas content might be the underlying
cause of the different gas density distributions and consequently,
enhanced HD/LD line luminosity ratios. 

Direct comparisons between high- and low-density molecular gas tracers 
to interpret the physical conditions of the molecular ISM in infrared 
luminous galaxies were also used in \citet{2008A&A...477..747B}. These 
authors propose that the gas density distribution reflects the evolutionary 
stage of a nuclear outburst in (U)LIRGs.  In their picture, the denser 
gas is depleted more rapidly than the lower density gas as the nuclear 
starburst progresses, and the molecular line ratios provide clues about a 
shift in the dominant heating source (UV versus X-ray photons).  

\citet{2008A&A...488L...5L} modeled some of the key molecular line ratios
presented in \citet{2008A&A...477..747B} and found that mechanical 
feedback heating is a crucial process to explain the low HNC(1-0)/HCN(1-0) 
ratio observed in some systems.  They provide additional support 
for a time-dependent model of the physical conditions in the nuclear
regions of luminous infrared galaxies, according to which the molecular 
ISM switches from a high-density ($\sim 10^5$~cm$^{-3}$) phase, dominated 
by stellar heating, to a lower-density ($\sim 10^{4.5}$~cm$^{-3}$) phase, 
dominated by mechanical heating.

We explore a varying HD/LD line luminosity ratio further in the next 
section, where we present simulations of individual disk galaxies as 
well as mergers of gas-rich galaxies.

\section{Model Line Ratios}
\label{sec:model}

In order to investigate whether X-ray induced chemistry is necessary
to drive the observed trends of line ratio with infrared luminosity
\citep[e.g.][]{2006ApJ...640L.135G}, we compare observed molecular 
line ratios with fixed-abundance numerical simulations 
\citep{2008ApJ...684..996N}. 

\begin{figure*}
\epsscale{0.85}
\plotone{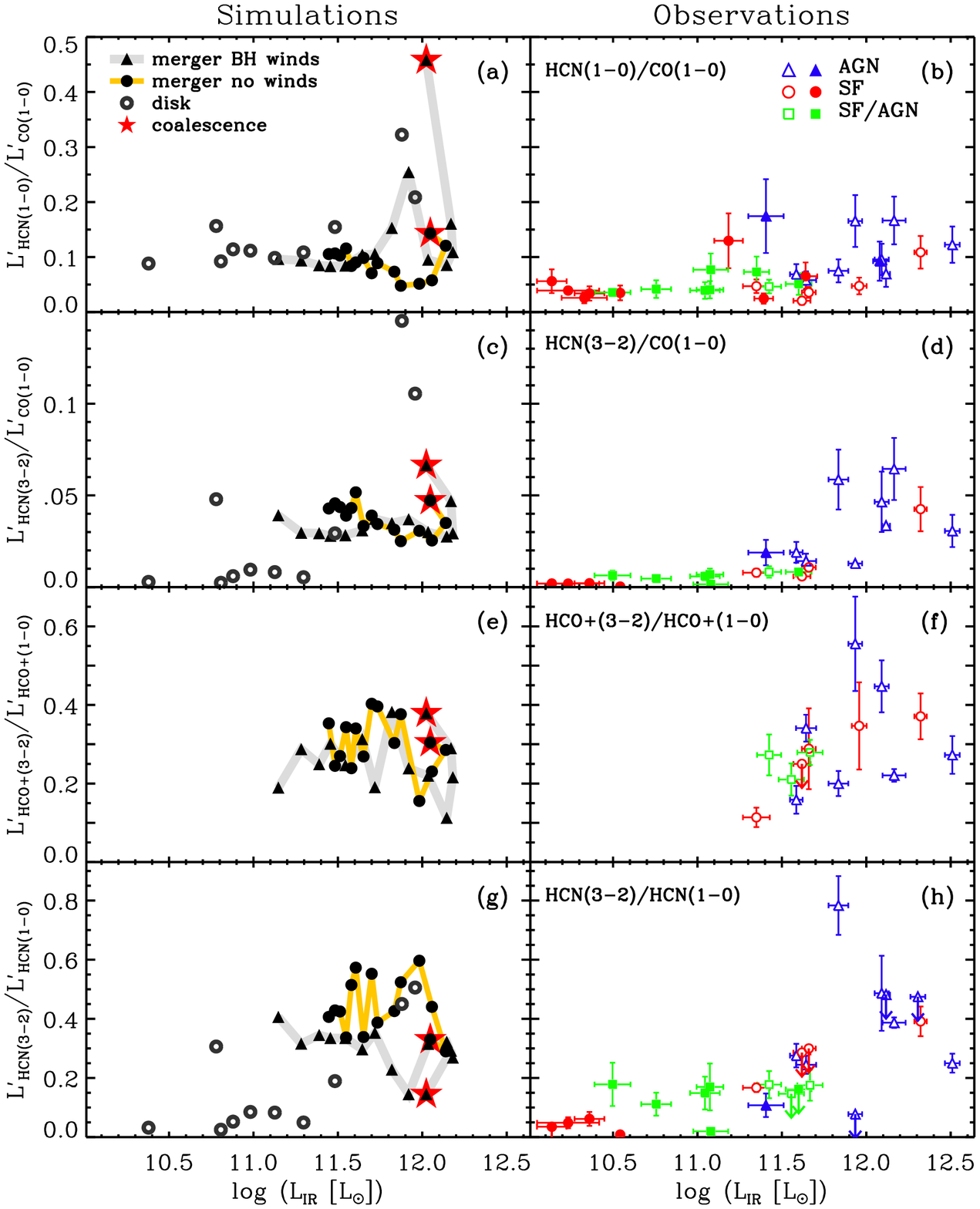}
\caption{Several molecular line luminosity ratios as a function of
  infrared luminosity.  The simulation results are shown in the
  left-hand column while the observations for the corresponding ratio
  are shown on the right.  From top to bottom, the luminosity ratios
  are: \hcnOne/\coOne ({\it a,b}), \hcnThr/\coOne ({\it c,d})
  , \hcoThr/\hcoOne ({\it e,f}), and \hcnThr/\hcnOne
  ({\it g,h}).  This figure shows evolutionary tracks of
  equal-mass merger simulations with (gray line) and without (yellow
  line) SMBH feedback combined with simulations of individual gas-rich
  disk galaxies (open circles). The tracks start at black-hole coalescence 
  (large star symbol), and each time step is marked with a black
  symbol.  Although their dynamical range differs slightly, the 
  models successfully reproduce the spread observed in the data.  
  Plotting symbols for the observations 
  are as follows: star-forming galaxies (SF) ({\em red circles}); 
  AGN ({\em blue triangles}); and SF/AGN ({\em green squares}). We use
  open symbols for galaxies that belong to the GC08 sample and filled
  symbols otherwise.
  (A color version of this figure is available in the online journal.)}
\label{fig:hcnco_model}
\end{figure*}

In the left column of Figure~\ref{fig:hcnco_model}, we show the model
predictions, and compare them directly to the observations in the
right column.  The HCN and CO simulations were taken from the study of
\citep{2008ApJ...684..996N}, while HCO$^+$ simulations of 
gas-rich galaxy mergers were run specifically for this work. When 
available, the isolated disk galaxies include a large dynamic range 
of gas fractions and masses (see \S~\ref{sec:sample}).  
The two galaxy merger simulations are identical in all ways except for 
their feedback implementation. The thick gray curve represents a galaxy merger
in which 0.5\% of the accreted mass energy onto the central black hole is
re-injected into the surrounding ISM as thermal energy input, while the
merger shown by the thin yellow curve does not include AGN feedback. 
Each time step of 5~$h^{-1}$~Myr is shown with a filled triangle 
(circle) for the model with(without) black-hole feedback. 
During their peak burst (when the galaxy may be most visible as a ULIRG),
AGN winds can vary both the star formation history and gas density 
profiles in the galaxies \citep[e.g. ][]{2005MNRAS.361..776S}. Hence, 
these two models effectively serve as two different galaxies probing 
different line ratio-$L_{\rm IR}$ relations. The infrared luminosity 
of the simulated galaxies is estimated from their known SFR.  We use the
\citet{1998ApJ...498..541K} conversion $L_{FIR} = (5.5 \times 10^{9}) \times SFR$
and our observed relation $L_{IR} = 1.38 \times L_{FIR}$ found in
Figure~\ref{fig:FIR_IR}.

The molecular line luminosity ratios considered here are \hcnOne/\coOne,
\hcnThr/\coOne, \hcoThr/\hcoOne\ and \hcnThr/\hcnOne.  As before, we
show the galaxies overlapping with the GC08 sample with open symbols,
whereas filled symbols are for observations from our primary data set
only.  Because galaxies with HCO$^{+}$ observations strictly belong to
 the GC08 sample, they occupy the brighter end of the $L_{IR}$ range, 
which is well-sampled by the merger models.  The isolated
gas-rich galaxies (open circles) cover the low-end of the infrared
luminosity, with the exception of two cases at $L_{IR} \sim 10^{11.5}
- 10^{12} L_{\sun}$, which were designed to explore extreme conditions
that are physically unrealistic at low-redshift (e.g. the most massive
$M_{\rm DM} \sim$10$^{13} M_\odot$ galaxies).

Generally speaking, there is an excellent correspondence between the
range of the observations and models in their molecular line ratios.  
At large infrared luminosities (SFRs), the line ratios tend to increase. 
This effect is simply a manifestation of the fact that the galaxies at 
the high infrared luminosity range are undergoing a starburst event. 
When the fraction of dense gas in a galaxy increases (for example, 
owing to a merger) so does the rate at which stars form. Large amounts 
of dense gas also imply that starbursting galaxies 
are more easily able to excite high critical density tracers (such as
various transitions of HCN or HCO$^+$), thus increasing the observed
HD/LD line ratios.  Because the simulations include constant, 
Galactic-based abundances without chemistry-driven variations, the
agreement between model and data demonstrate that chemistry-driven
abundance variations are not necessary to produce the observed line 
ratios.

We note that generally the dynamic range of the modeled line ratios
seems to match the observations reasonably well. However, we urge 
caution with a detailed comparison of the models and the data.
First, the simulations were designed to probe a large parameter space 
in gas fractions and galaxy masses. Some galaxies were specifically 
designed to probe relatively extreme conditions (e.g. initial gas fractions
$f_{\rm g}$=0.8, $M_{\rm DM} \sim$ 10$^{13} M_\odot$). Consequently,
individual model galaxies can be caught during a brief snapshot with
extreme line ratios, and may not exactly map to a particular galaxy
from observed galaxy samples in Figure~\ref{fig:hcnco_model}. 
Second, the galaxies in the models
were not chosen to precisely mimic the relative number of isolated
galaxies versus mergers in the GS04a,b samples. Thus, the clustering
of simulated points in Figure~\ref{fig:hcnco_model} may not
exactly match those in the observations. Third, the $L_{\rm IR}$ in
the models was calculated using a linear mapping from the SFR. While
we do not consider AGN contribution to $L_{IR}$ in the simulations, there is
some in the observations.  This is likely to be the reason why the
observations extend to brighter $L_{IR}$ ($> 10^{12.5}~L_{\sun}$) 
compared to the simulations, which seem to reach a ceiling at $L_{IR} 
\sim 10^{12.2}~L_{\sun}$.

\section{Possible Chemistry Effects}
\label{sec:Chem}

It has been suggested that the radiation field associated with an AGN
influences the abundance of HCN with respect to other molecules.  For
example, emitted X-rays could cause X-ray dominated regions (XDRs)
which have different properties than regular photon dominated regions
(PDRs) found around star-forming regions. Previous authors claim that 
conditions existing in these XDRs affect molecular gas abundance 
\citep{2006ApJ...646L..37L, 2007A&A...464..193A, 2007A&A...468L..63K}.

On the other hand, \citet{2008A&A...477..747B} used multiple molecular line
ratios to distinguish between XDR and PDR conditions, and found that
most of the (U)LIRGs in their sample are dominated by PDRs.  They used 
tracers sensitive to column density ($N_H$) to distinguish between an
elevated HCN/CO intensity ratio resulting from a high-$N_H$ PDR or from 
a low-$N_H$ XDR, and found the former to be more likely.

\begin{figure*}
\centering
\plotone{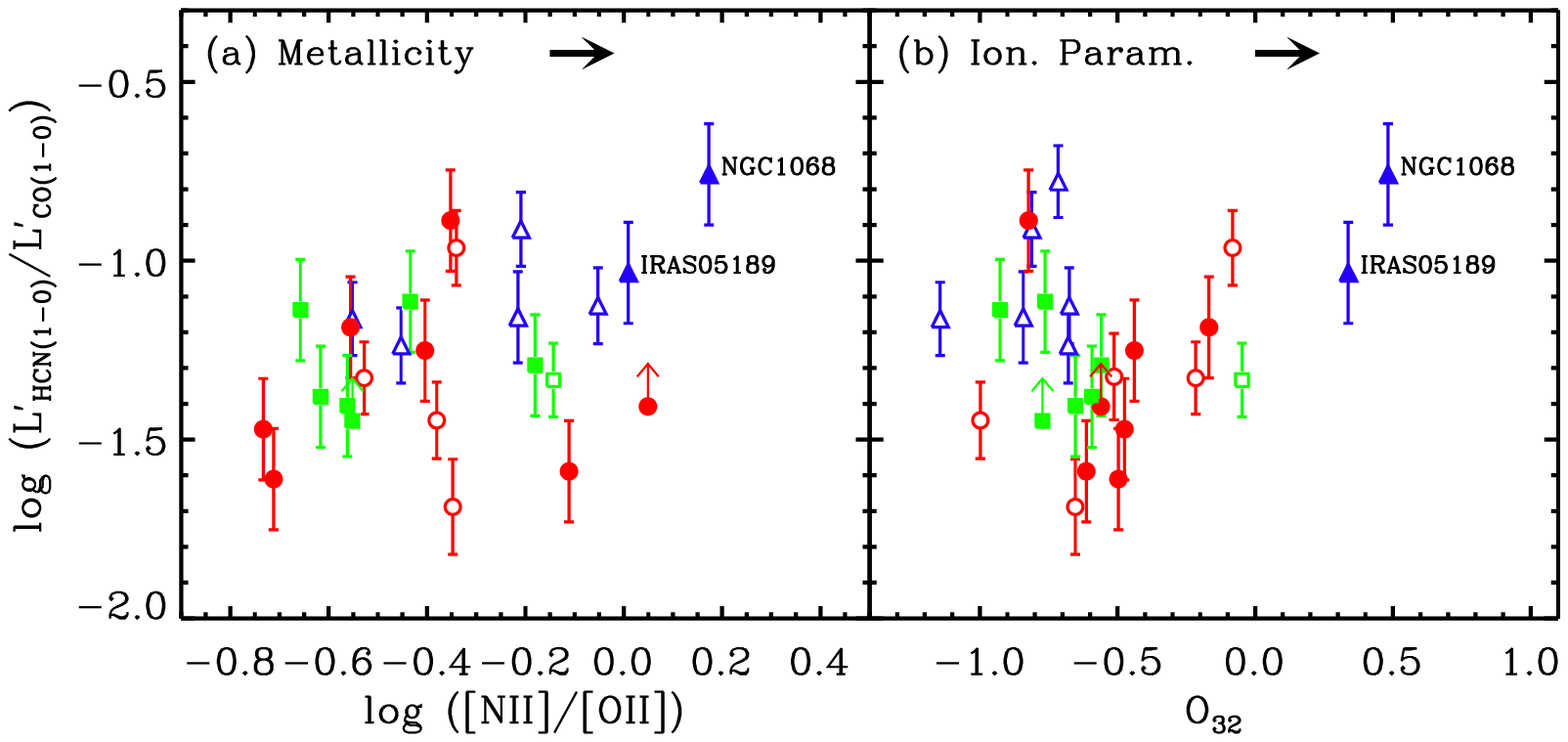}
\caption{\hcnOne/\coOne\ luminosity ratio as a function of nebular emission-line 
  diagnostics of the diffuse ISM metallicity ({\it Left}) and ionization 
  parameter ({\it Right}).  There is no obvious trend except for two outliers 
  for which the high \hcnOne/\coOne\ luminosity ratio could potentially
  be associated with enhanced metallicity and/or ionization in the ISM.
  Plotting symbols represent the optical spectral type: star-forming
  (SF) ({\em red circles}); AGN ({\em blue triangles}); and SF/AGN
  ({\em green squares}). We use open symbols for galaxies that belong to 
  the GC08 sample and filled symbols otherwise.
  (A color version of this figure is available in the online journal.)}
\label{fig:hcnco_Z}
\end{figure*}

Below we present optical spectral diagnostics associated with
ISM metallicity and ionization parameter (Figure~\ref{fig:hcnco_Z}).
The emission-line ratio \niilam/\oiilam\ has been shown to correlate
with the gas-phase oxygen abundance $12+$log$(O/H)$ while being less
affected by the presence of an AGN than other metallicity diagnostics
\citep{2002ApJS..142...35K, 2008ApJ...681.1183K}. Meanwhile, the emission-line
ratio $O_{32} \equiv$ log$(\oiiilam/\oiilam)$ traces the ionization
parameter.  The correspondence between the observed ratio and the
ionization depends slightly on the metallicity so the two panels of 
Figure~\ref{fig:hcnco_Z} should be interpreted in conjunction.

Most galaxies are scattered without obvious trends in the
\hcnOne/\coOne-$\niilam/\oiilam$ and \hcnOne/\coOne-$O_{32}$ planes.
The correlations are very weak or non-existent, especially if we
exclude the two outliers (IRAS~05189-2524 and the more extreme NGC~1068).
In the first panel, the correlation coefficient $r=0.42$ drops to
$r=0.24$ if we exclude NGC~1068 and down further to $r=0.15$ if we
exclude both outliers mentioned above.  In the second panel, the
correlation coefficient drops from $r=0.26$ to $r=-0.09$ if we exclude
both outliers.  NGC~1068 (and possibly IRAS~05189-2524) may be a 
special case where unusual ionization or chemistry effects could play 
a significant role in producing the observed \hcnOne/\coOne\ luminosity
ratios.

Our result is in agreement with findings of \citet{2004A&A...419..897U} who 
conclude that the circumnuclear region of NGC~1068 is effectively a 
giant XDR. These authors used a combination of single dish and interferometry 
data of several molecular species to rule out alternative explanations 
for the observed high HCN/CO ratio \citep{1994ApJ...426L..77T}.
NGC~1068 is also observed to have the largest MIR-excess (it lies 
$4.5\sigma$ from the mean on Figure~\ref{fig:FIR_IR}), a regime where
MIR-pumping can promote \hcnOne\ emission by exciting a 
bending mode at $14~\mu$m \citep{1995A&A...300..369A}.

While the optical emission line ratios used here may give a good 
indication of the metallicity and ionization state of the diffuse ISM, 
it is not clear how representative they are of the conditions present 
in denser molecular gas.  Therefore, it may not be surprising that 
we do not see obvious trends in either panel.  This caveat is 
especially important if young stars (starburst) are the main source of 
ionization because UV photons do not penetrate deeply into the dense 
molecular.  On the other hand,
X-ray photons can penetrate through much larger column densities
of material, so it is plausible that the extreme conditions triggered 
by an AGN would occur over a large enough volume to effect 
both the diffuse and dense gas phases.  
If this were the case, we would expect a larger \hcnOne/\coOne\ 
luminosity ratio to couple with larger values of $O_{32}$
because an increased ionization means more free electrons, which
can accelerate the production of HCN molecules (as discussed in 
\S\ref{sec:intro}).  

Overall, our results indicate that cases with a genuine abundance 
change in HCN may exist but are the exception rather than the rule.

\section{Conclusions}
\label{sec:conclusions}

Using a sample of 34 nearby infrared luminous galaxies 
($10^{10} < L_{IR} < 10^{12.5}~L_{\sun}$), we characterized 
the infrared luminosity dependence of various molecular gas tracers. 

\begin{itemize}

\item{The presence of AGN was assessed using the optical {\it BPT} diagram.
In agreement with previous publications, we find a more frequent occurrence 
of AGN in more IR-luminous galaxies. This may be related to the 
availability of larger amounts of dense gas during the mergers of 
gas-rich galaxies.
}

\item{The molecular transitions used, \coOne, \hcoOne, \hcnOne, \hcoThr\ 
and \hcnThr, span 4 orders of magnitude in critical density.  We find that
the relationship between (F)IR luminosity and molecular line luminosity
$L_{mol}$ is shallower for transitions with higher $n_{crit}$.  This 
trend is in agreement with theoretical models of N08 and KT07 and can 
be explained by the varying degree of thermalization of the gas giving
rise to molecular line emission.
}

\item{Trends of molecular line ratios with $L_{IR}$ are consistent 
with an increased molecular gas density in more IR-bright galaxies. 
This result agrees with the picture presented by 
\citet{2004ApJ...606..271G} and \citet{2005ApJ...635L.173W}.
When comparing high-density and low-density molecular gas tracers, 
we observe an increase in their luminosity ratio ($L_{HD}/L_{LD}$)
with increasing infrared luminosity.  Interestingly, this trend 
vanishes when comparing two tracers with nearly equal critical 
density (\hcnOne and \hcoThr).  We infer that the main driver of 
the enhanced HD/LD luminosity ratio is the molecular gas density 
distribution, in agreement with the observations that ULIRGs host 
large reservoirs of dense molecular gas in their central regions.
}

\item{We compare our observed molecular line ratios with theoretical values 
obtained from a set of galaxy models.  We consider SPH simulations
of two galaxy mergers as well as individual gas-rich galaxies.  With 
constant Galactic abundances, the models successfully produce an enhanced
HD/LD luminosity ratios at brighter infrared luminosity.  This 
provides additional support for a higher molecular gas density in 
galaxies that are extremely gas-rich or undergoing gas-rich mergers. 
Indeed, this result suggests that AGN-induced chemistry
(or other) effects may not be necessary to reproduce the observations, 
but rather that AGNs are more likely to reside in galaxies that 
are very gas-rich and/or experiencing a merger.
The simulations also demonstrate important variations in molecular
luminosity ratio with the evolutionary stage of the mergers. 
These variations may be connected with the dynamics of the molecular 
gas during mergers (inflows, outflows, gas compression, etc.).
}

\item{We investigate possible chemistry effects on the well-known 
\hcnOne/\coOne\ luminosity ratio using common optical emission line diagnostics 
of the ISM ionization parameter and metallicity. With the exception of 
one of two outliers, we do not observe significant correlations between
the molecular line ratio and these properties.  Decoupling between 
the dense molecular gas traced by \coOne\ and \hcnOne\ and the more 
diffuse gas that traces the optical nebular line may be responsible 
for the absence of correlation.  However, we note that NGC~1068 shows 
extreme conditions in the sense of having the largest optically-measured 
ionization parameter as well as the largest MIR-excess.  This outlier
is thus more subject to chemistry and/or MIR-pumping effects quoted 
in the literature.  We emphasize that NGC~1068 has properties that are
very distinct from the rest of our sample, and thus may not be 
representative of its class.
}

\item{More high-density molecular line observations would be extremely
beneficial to confirm the trends outlined in this work. Only half of our
combined sample of 34 galaxies have data for all five molecular lines 
used here. In particular, future observations should target galaxies 
with fainter IR luminosities to test whether the results presented here 
extend into the lower luminosity regime.
Higher-resolution studies of high-density tracers may complement this work 
by allowing for a more detailed analysis of the density and excitation 
structure of the molecular gas within galaxies. Such studies could provide 
information on the \emph{local} influence of AGN, whereas in this work 
we probe the \emph{global} influence of AGN.
}

\item{Overall, our results support that \hcnOne\ is a valid tracer 
of dense molecular gas in galaxies even in the presence of AGN.
We expect scatter in the relationship due to variations in temperature 
and possible radiative (de-)excitation of this transition.  Also,
the dense-to-total molecular gas fraction is expected to differ from galaxy
to galaxy, especially in systems undergoing significant mergers.
}

\end{itemize}

\begin{acknowledgements}
The authors wish to thank Yu Gao and Neal Evans for useful comments 
and Leo Blitz for sharing early results about molecular outflows.  
This work also benefited from invaluable input from Willem Baan 
\& Javier Grac\'{i}a-Carpio.
We are grateful to the anonymous referee for a constructive report 
which helped improve this manuscript.  S. Juneau acknowledges Mark 
Dickinson for his support and advice.  S. Juneau was partially funded 
by a FQRNT doctoral fellowship (Qu\'{e}bec, Canada).

This research has made use of the NASA/IPAC Extragalactic Database 
(NED) which is operated by the Jet Propulsion Laboratory, California 
Institute of Technology, under contract with the National Aeronautics 
and Space Administration. 
\end{acknowledgements}

\bibliographystyle{apj}
\bibliography{hcn_ulirg}


\begin{deluxetable*}{lccccccccc}
\tabletypesize{\scriptsize}
\tablecolumns{10}
\tablewidth{550pt}
\tablecaption{Infrared, far-infrared and molecular line luminosities
  for our combined sample of 34 galaxies.  The optical classification
  is included in the last column.}  \tablehead{ \colhead{Source} &
  \colhead{$D_{L}$} & \colhead{$L_{IR}$} & \colhead{$L_{FIR}$} &
  \colhead{\LpcoOne}\tablenotemark{a} & \colhead{\LphcnOne}\tablenotemark{a} & \colhead{\LphcnThr} &
  \colhead{\LphcoOne} & \colhead{\LphcoThr} & \colhead{Class}
  \\ \colhead{ } & \colhead{(Mpc)} & \colhead{log($L_{\sun}$)} &
  \colhead{log($L_{\sun}$)} & \colhead{log($L'$)} &
  \colhead{log($L'$)} & \colhead{log($L'$)} & \colhead{log($L'$)} &
  \colhead{log($L'$)} & \colhead{ } }
\startdata 
       IC 1623  &    84 $\pm$  6  &  11.64 $\pm$ 0.07  &  11.48 $\pm$ 0.07  &  10.14 $\pm$ 0.10  &   8.95 $\pm$ 0.10  &       \dots        &       \dots        &       \dots      &  SF  \\
      NGC  520  &    37 $\pm$  3  &  11.04 $\pm$ 0.09  &  10.93 $\pm$ 0.09  &   9.38 $\pm$ 0.10  &   7.97 $\pm$ 0.10  &   7.15 $\pm$ 0.09  &       \dots        &       \dots      &  SF/AGN  \\
      NGC  660  &    13 $\pm$  1  &  10.50 $\pm$ 0.11  &  10.37 $\pm$ 0.11  &   8.80 $\pm$ 0.10  &     $>$ 7.35       &   6.60 $\pm$ 0.11  &       \dots        &       \dots      &  SF/AGN  \\
      NGC  695  &   135 $\pm$  7  &  11.62 $\pm$ 0.05  &  11.48 $\pm$ 0.05  &   9.98 $\pm$ 0.10  &   8.29 $\pm$ 0.09  &     $<$ 7.75       &   8.44 $\pm$ 0.06  &     $<$ 7.84     &  SF  \\
      NGC 1068  &    16 $\pm$  1  &  11.41 $\pm$ 0.10  &  11.04 $\pm$ 0.10  &   9.31 $\pm$ 0.10  &   8.55 $\pm$ 0.10  &   7.58 $\pm$ 0.09  &       \dots        &       \dots      &  AGN      \\
      NGC 1144  &   119 $\pm$  7  &  11.39 $\pm$ 0.06  &  11.27 $\pm$ 0.06  &  10.06 $\pm$ 0.10  &   8.45 $\pm$ 0.10  &       \dots        &       \dots        &       \dots      &  SF  \\
      NGC 1614  &    67 $\pm$  5  &  11.60 $\pm$ 0.08  &  11.40 $\pm$ 0.08  &   9.45 $\pm$ 0.10  &   8.15 $\pm$ 0.10  &     $<$ 7.36       &       \dots        &       \dots      &  SF/AGN  \\
05189-2524\tablenotemark{b}  &   180 $\pm$  7  &  12.08 $\pm$ 0.04  &  11.84 $\pm$ 0.04  &   9.87 $\pm$ 0.10  &   8.84 $\pm$ 0.10  &       \dots      &       \dots      &       \dots      &  AGN      \\
      NGC 2146  &    17 $\pm$  2  &  11.08 $\pm$ 0.10  &  10.93 $\pm$ 0.10  &   9.24 $\pm$ 0.10  &   8.12 $\pm$ 0.10  &   6.41 $\pm$ 0.10  &       \dots        &       \dots      &  SF/AGN  \\
        Arp 55  &   170 $\pm$  7  &  11.66 $\pm$ 0.04  &  11.56 $\pm$ 0.04  &  10.14 $\pm$ 0.10  &   8.69 $\pm$ 0.04  &     $<$ 8.17       &   8.65 $\pm$ 0.06  & 8.11 $\pm$ 0.12  &  SF  \\
      NGC 2903  &     8 $\pm$  1  &  10.23 $\pm$ 0.18  &  10.08 $\pm$ 0.18  &   8.68 $\pm$ 0.10  &     $>$ 7.27       &   5.95 $\pm$ 0.10  &       \dots        &       \dots      &  SF  \\
     UGC 05101  &   170 $\pm$  7  &  11.93 $\pm$ 0.04  &  11.84 $\pm$ 0.04  &   9.76 $\pm$ 0.10  &   8.98 $\pm$ 0.04  &     $<$ 7.87       &   8.70 $\pm$ 0.07  & 8.45 $\pm$ 0.04  &  AGN      \\
      NGC 3079  &    19 $\pm$  1  &  10.76 $\pm$ 0.09  &  10.65 $\pm$ 0.09  &   9.54 $\pm$ 0.10  &   8.16 $\pm$ 0.10  &   7.20 $\pm$ 0.08  &       \dots        &       \dots      &  SF/AGN  \\
      NGC 3628  &    11 $\pm$  1  &  10.36 $\pm$ 0.09  &  10.24 $\pm$ 0.09  &   9.24 $\pm$ 0.10  &   7.77 $\pm$ 0.10  &   6.56 $\pm$ 0.10  &       \dots        &       \dots      &  SF  \\
      NGC 3893  &    15 $\pm$  1  &  10.14 $\pm$ 0.09  &  10.00 $\pm$ 0.09  &   8.71 $\pm$ 0.10  &   7.46 $\pm$ 0.10  &     $<$ 6.00       &       \dots      &       \dots      &  SF  \\
      NGC 4414  &    17 $\pm$  0  &  10.54 $\pm$ 0.02  &  10.40 $\pm$ 0.02  &   9.22 $\pm$ 0.10  &   7.76 $\pm$ 0.10  &     $<$ 5.70       &       \dots      &       \dots      &  SF  \\
       Arp 193  &   106 $\pm$  7  &  11.64 $\pm$ 0.06  &  11.55 $\pm$ 0.06  &   9.72 $\pm$ 0.10  &   8.49 $\pm$ 0.03  &   7.88 $\pm$ 0.04\tablenotemark{d}  &   8.70 $\pm$ 0.03  &   8.23 $\pm$ 0.03  &  AGN      \\
      NGC 5194  &     8 $\pm$  1  &  10.33 $\pm$ 0.13  &  10.20 $\pm$ 0.13  &   9.13 $\pm$ 0.10  &   7.54 $\pm$ 0.10  &       \dots        &       \dots      &       \dots      &  SF  \\
     UGC 08696  &   165 $\pm$  7  &  12.09 $\pm$ 0.04  &  11.98 $\pm$ 0.04  &   9.89 $\pm$ 0.10  &   8.87 $\pm$ 0.05  &   8.56 $\pm$ 0.09\tablenotemark{e}  &   8.88 $\pm$ 0.05  &   8.53 $\pm$ 0.04  &  AGN      \\
       Arp 220  &    85 $\pm$  6  &  12.16 $\pm$ 0.07  &  12.08 $\pm$ 0.07  &  10.01 $\pm$ 0.10  &   9.23 $\pm$ 0.01  &   8.82 $\pm$ 0.02\tablenotemark{d}  &   8.89 $\pm$ 0.02  &   8.23 $\pm$ 0.02  &  AGN      \\
      NGC 6240  &   111 $\pm$  7  &  11.84 $\pm$ 0.06  &  11.69 $\pm$ 0.06  &  10.01 $\pm$ 0.10  &   8.89 $\pm$ 0.04  &   8.78 $\pm$ 0.04\tablenotemark{e}  &   9.15 $\pm$ 0.02  &   8.45 $\pm$ 0.06  &  AGN      \\
17208-0014\tablenotemark{b}  &   187 $\pm$  7  &  12.32 $\pm$ 0.04  &  12.25 $\pm$ 0.04  &  10.24 $\pm$ 0.10  &   9.27 $\pm$ 0.03  &   8.87 $\pm$ 0.04\tablenotemark{e}  &   9.10 $\pm$ 0.04  &   8.67 $\pm$ 0.05  &  SF  \\
      NGC 6701  &    61 $\pm$  5  &  11.07 $\pm$ 0.08  &  10.94 $\pm$ 0.08  &   9.59 $\pm$ 0.10  &   8.20 $\pm$ 0.10  &   7.43 $\pm$ 0.13  &       \dots      &       \dots      &  SF/AGN  \\
      NGC 7130  &    71 $\pm$  6  &  11.35 $\pm$ 0.07  &  11.22 $\pm$ 0.07  &   9.73 $\pm$ 0.10  &   8.59 $\pm$ 0.10  &       \dots      &       \dots      &       \dots      &  SF/AGN  \\
       IC 5179  &    50 $\pm$  4  &  11.18 $\pm$ 0.08  &  11.05 $\pm$ 0.08  &   9.49 $\pm$ 0.10  &   8.61 $\pm$ 0.10  &       \dots      &       \dots      &       \dots      &  SF  \\
      NGC 7469  &    69 $\pm$  2  &  11.59 $\pm$ 0.04  &  11.39 $\pm$ 0.04  &   9.59 $\pm$ 0.10  &   8.43 $\pm$ 0.02  &   7.87 $\pm$ 0.05\tablenotemark{d}  &   8.59 $\pm$ 0.01  &   7.79 $\pm$ 0.09  &  AGN      \\
23365+3604\tablenotemark{b}  &   269 $\pm$  6  &  12.12 $\pm$ 0.02  &  11.97 $\pm$ 0.02  &   9.94 $\pm$ 0.10  &   8.78 $\pm$ 0.08  &     $<$ 8.47\tablenotemark{f}     &   8.61 $\pm$ 0.10  &       \dots      &  AGN      \\
      NGC 7771  &    60 $\pm$  5  &  11.35 $\pm$ 0.08  &  11.23 $\pm$ 0.08  &   9.96 $\pm$ 0.10  &   8.63 $\pm$ 0.01  &   7.85 $\pm$ 0.03\tablenotemark{d}  &   8.62 $\pm$ 0.03  &   7.67 $\pm$ 0.08  &  SF  \\
       Mrk 331  &    77 $\pm$  6  &  11.42 $\pm$ 0.07  &  11.28 $\pm$ 0.07  &   9.75 $\pm$ 0.10  &   8.41 $\pm$ 0.02  &   7.66 $\pm$ 0.10\tablenotemark{e}  &   8.50 $\pm$ 0.02  &   7.94 $\pm$ 0.07  &  SF/AGN  \\
       Mrk 231\tablenotemark{c}  &   188         &  12.51 $\pm$ 0.07  &  12.26 $\pm$ 0.07  &  10.00 $\pm$ 0.10  &   9.09 $\pm$ 0.03  &   8.49 $\pm$ 0.05\tablenotemark{e}  &   9.07 $\pm$ 0.04  &   8.50 $\pm$ 0.06  &  AGN      \\
12112+0305\tablenotemark{b,c}  &   335         &  12.31 $\pm$ 0.07  &  12.19 $\pm$ 0.07  &       \dots        &   9.22 $\pm$ 0.09  &     $<$ 8.90\tablenotemark{e}     &   8.99 $\pm$ 0.11  &       \dots      &  AGN      \\
     VII Zw 31\tablenotemark{c}  &   241         &  11.96 $\pm$ 0.07  &  11.83 $\pm$ 0.07  &  10.16 $\pm$ 0.10  &   8.84 $\pm$ 0.07  &     $<$ 8.57      &   8.90 $\pm$ 0.06  &   8.44 $\pm$ 0.10  &  SF  \\
      Arp 299A\tablenotemark{c}  &    50 $\pm$  4  &  11.67 $\pm$ 0.07  &  11.48 $\pm$ 0.07  &       \dots        &   8.21 $\pm$ 0.02  &   7.45 $\pm$ 0.11\tablenotemark{e}  &   8.48 $\pm$ 0.02  &   7.93 $\pm$ 0.04  &  SF/AGN  \\
      Arp 299B\tablenotemark{c}  &    50 $\pm$  4  &  11.56 $\pm$ 0.07  &  11.29 $\pm$ 0.07  &       \dots        &   8.02 $\pm$ 0.03  &     $<$ 7.19\tablenotemark{e}     &   8.22 $\pm$ 0.03  &   7.55 $\pm$ 0.07  &  SF/AGN  \\
\enddata
\tablenotetext{a}{~ When no formal uncertainties are available for 
   published CO(1-0) or HCN(1-0) line luminosities, we estimate the 
   calibration error to be 0.1~dex (20$-$25\%).}
\tablenotetext{b}{~ Full galaxy name starts with {\it IRAS}.}
\tablenotetext{c}{~ These five galaxies are not part of our primary 
   sample. $L_{IR}$ and $L_{FIR}$ were obtained from GC08 after converting
   their luminosity distance to our assumed cosmology (\{$\Omega_{M}, 
   \Omega_{\Lambda}$, H$_0$\} = \{0.3, 0.7, 70~km s$^{-1}$ Mpc$^{-1}$\}).
   Because GC08 do not provide errors on $L_{IR}$ and $L_{FIR}$, we use 
   the mean error for the rest of our sample ($\approx$ 0.07~dex).}
\tablenotetext{d}{~ Average of B08 and GC08 values.}
\tablenotetext{e}{~ Measurement from GC08.}
\tablenotetext{f}{~ Upper limit in both B08 and GC08 surveys.}
\label{tab:data}
\end{deluxetable*}

\begin{deluxetable}{lccccc}
\tablecolumns{6}
\centering
\tablecaption{Slopes of log($L_{IR}$) $-$ log($L'_{mol}$) and log($L_{FIR}$) $-$ log($L'_{mol}$) for five molecular lines.
The transitions are ordered in increasing critical density ($n_{crit}$). For each transition, we report the difference
between the two values of slopes ($\beta_{FIR}$-$\beta_{IR}$) and we identify the sample used for the calculations.}
\tablehead{
\colhead{Transition} & \colhead{log($n_{crit} [cm^{-3}]$)} & \colhead{$\beta_{IR}$} & \colhead{$\beta_{FIR}$} & \colhead{$\beta_{FIR}$-$\beta_{IR}$} & \colhead{Sample}
}
\startdata
CO($J=1-0$)         & 3.3  & 1.30  $\pm$ 0.17  &  1.34 $\pm$  0.16 & 0.04 & combined \\
HCO$^{+}$($J=1-0$)  & 5.3  & 0.99  $\pm$ 0.26  &  1.07 $\pm$  0.26 & 0.08 & GC08 \\
HCN($J=1-0$)        & 6.5  & 1.13  $\pm$ 0.11  &  1.14 $\pm$  0.10 & 0.01 & combined \\
HCO$^{+}$($J=3-2$)  & 6.6  & 0.81  $\pm$ 0.21  &  0.90 $\pm$  0.19 & 0.09 & GC08 \\
HCN($J=3-2$)        & 7.7  & 0.70  $\pm$ 0.09  &  0.71 $\pm$  0.08 & 0.01 & combined \\
\enddata
\label{tab:slopes}
\end{deluxetable}

\begin{deluxetable}{lccccc}
\tablecolumns{6}
\tablecaption{Slopes of log($L'_{HD}$/$L'_{LD}$) $-$ log($L_{IR}$) and log($L'_{HD}$/$L'_{LD}$) $-$ log($L_{FIR}$)
for ten molecular line luminosity ratios.  For each line ratio, we report the contrast in critical density 
($R_{crit}$), the mean and standard deviations of the slope distributions ($\alpha_{IR}$ \& $\alpha_{FIR}$),
and the linear correlation coefficients ($r_{IR}$ \& $r_{FIR}$).}
\tablehead{
\colhead{Line Ratio} & \colhead{$R_{crit}$} & \colhead{$\alpha_{IR}$} & \colhead{$\alpha_{FIR}$} & \colhead{$r_{IR}$} & \colhead{$r_{FIR}$}
}
\startdata
HCO$^+_{3-2}$/HCN$_{1-0}$   & 0.1  &  -0.054 $\pm$ 0.24   & -0.068 $\pm$ 0.22  & -0.05 & -0.08 \\
HCN$_{1-0}$/HCO$^+_{1-0}$   & 1.1  &   0.32  $\pm$ 0.15   &  0.36  $\pm$ 0.13  & 0.60  &  0.66 \\
HCN$_{3-2}$/HCO$^+_{3-2}$   & 1.2  &   0.32  $\pm$ 0.39   &  0.38  $\pm$ 0.38  & 0.37  &  0.42 \\
HCO$^+_{3-2}$/HCO$^+_{1-0}$ & 1.3  &   0.24  $\pm$ 0.17   &  0.26  $\pm$ 0.15  & 0.48  &  0.53 \\
HCN$_{3-2}$/HCN$_{1-0}$     & 1.3  &   0.41  $\pm$ 0.11   &  0.41  $\pm$ 0.12  & 0.70  &  0.72 \\
HCO$^+_{1-0}$/CO$_{1-0}$    & 2.0  &   0.24  $\pm$ 0.20   &  0.22  $\pm$ 0.19  & 0.40  &  0.28 \\
HCN$_{3-2}$/HCO$^+_{1-0}$   & 2.4  &   0.53  $\pm$ 0.31   &  0.62  $\pm$ 0.30  & 0.68  &  0.75 \\
HCN$_{1-0}$/CO$_{1-0}$      & 3.1  &   0.23  $\pm$ 0.064  &  0.22  $\pm$ 0.065 & 0.53  &  0.49 \\
HCO$^+_{3-2}$/CO$_{1-0}$    & 3.2  &   0.48  $\pm$ 0.27   &  0.48  $\pm$ 0.26  & 0.60  &  0.60 \\
HCN$_{3-2}$/CO$_{1-0}$      & 4.4  &   0.70  $\pm$ 0.12   &  0.68  $\pm$ 0.11  & 0.87  &  0.87 \\
\enddata
\label{tab:ratios}
\end{deluxetable}

\clearpage

\appendix
\section{Observed HCN(3-2) Spectra}
\label{sec:append}

The Heinrich Hertz Submillimeter Telescope (HHT) \hcnThr\ spectra from the survey presented in \citet{2008ApJ...681L..73B} are displayed in Figure~\ref{fig:data}.

\begin{figure*}[hb]
\includegraphics[clip=true,width=\linewidth]{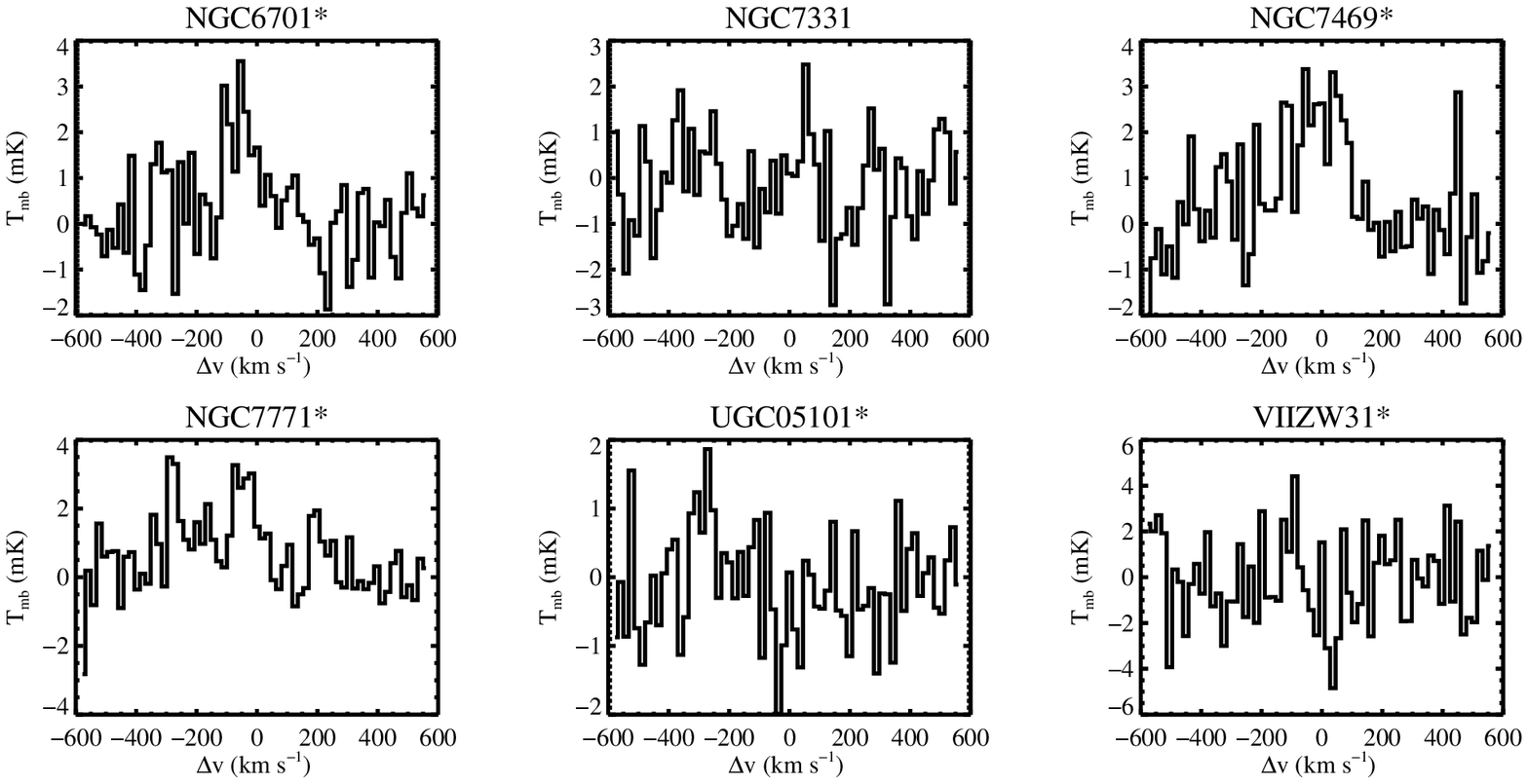}
\caption{HCN(3-2) spectra for all galaxies observed in B08.  The 23 galaxies 
overlapping with this sample are marked with an asterisk.  One galaxy was 
mapped (NGC~253, plus symbol).  These observations were combined with the 
HCN(3-2) data presented in \citet{2008A&A...479..703G} as described in 
\S\ref{sec:sample}.}
\label{fig:data}
\end{figure*}

\begin{figure*}
\addtocounter{figure}{-1}
\includegraphics[clip=true,width=\linewidth]{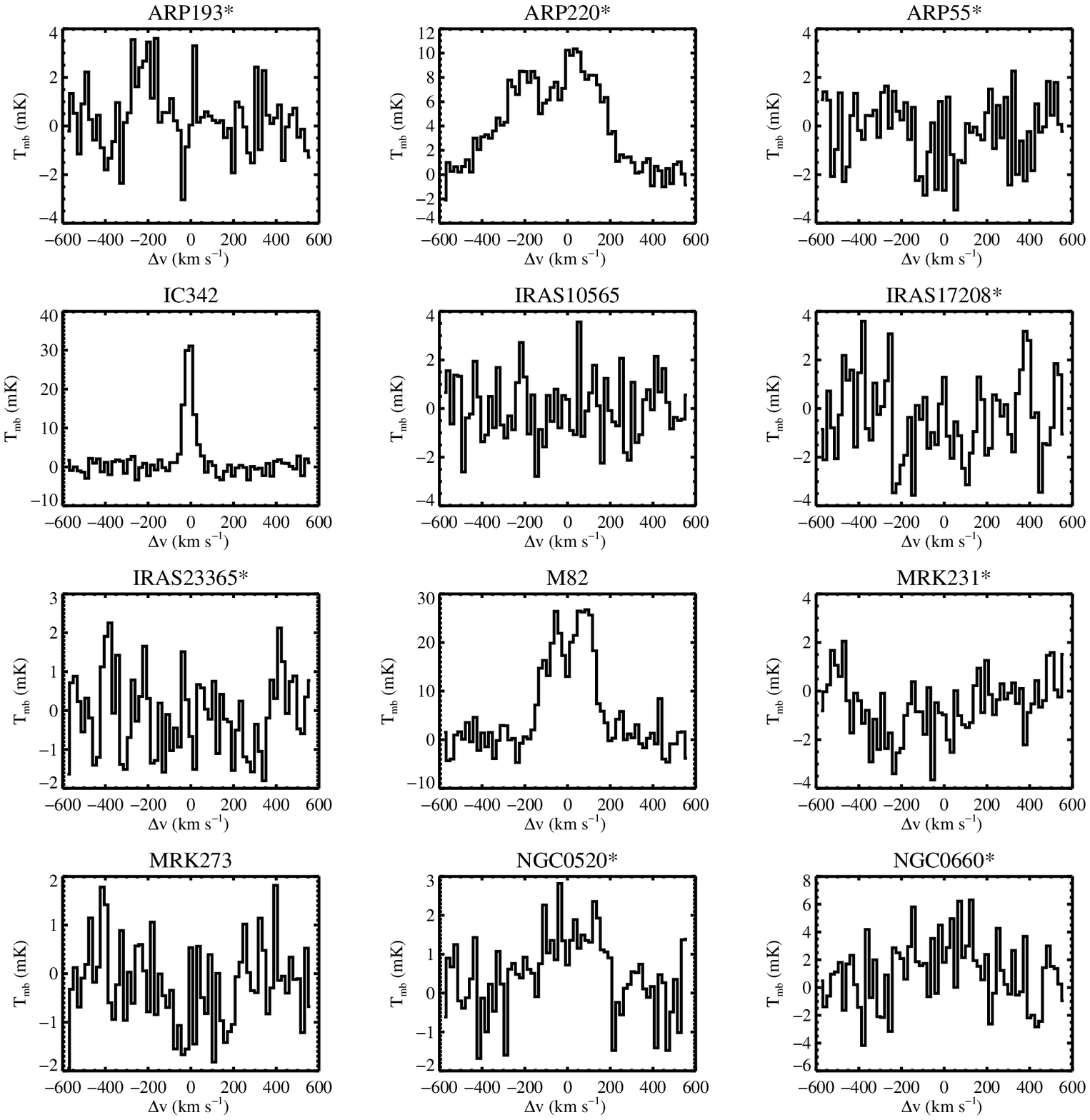}
\caption{Continued.}
\end{figure*}

\begin{figure*}
\addtocounter{figure}{-1}
\includegraphics[clip=true,width=\linewidth]{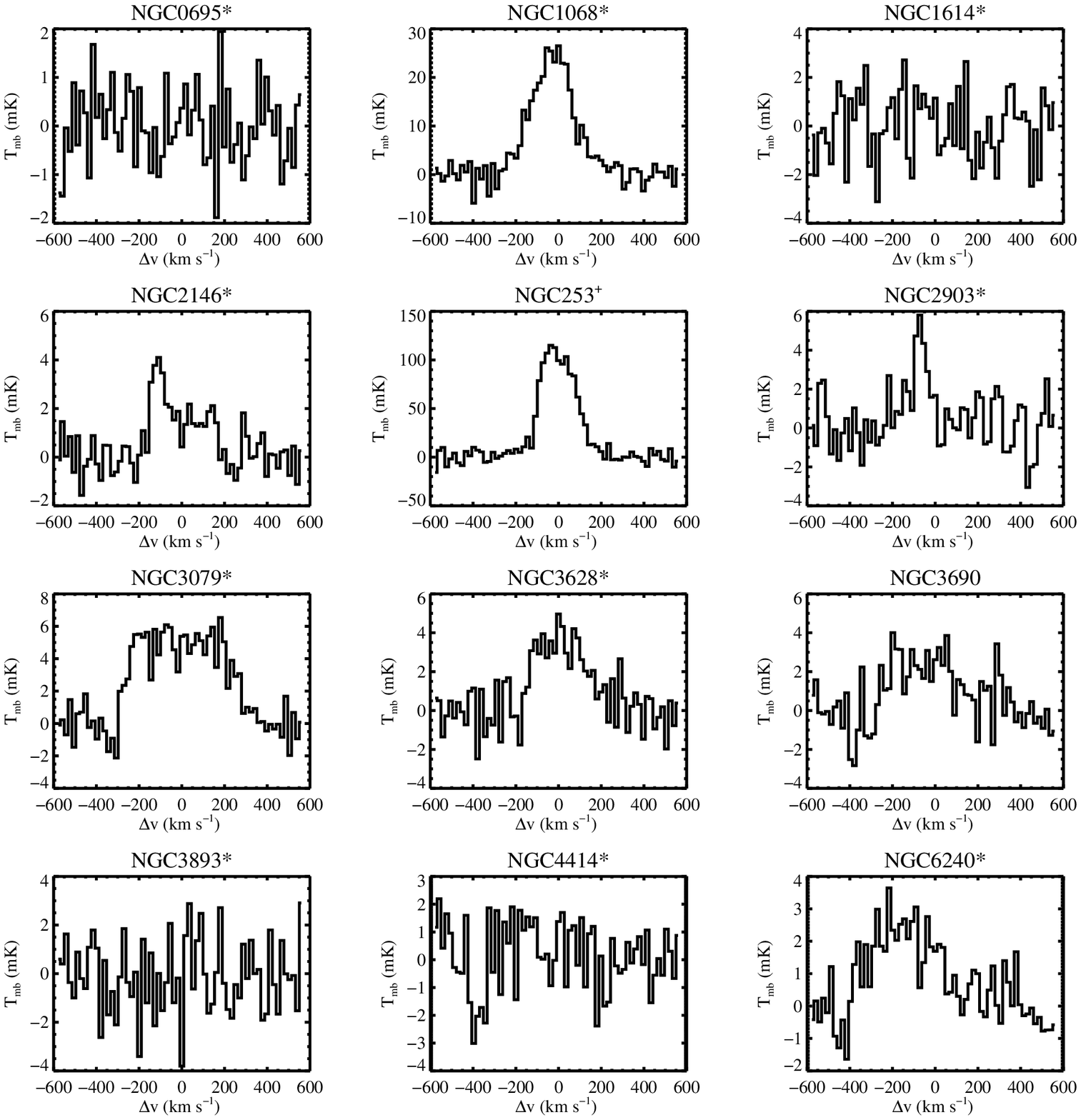}
\caption{Continued.}
\end{figure*}


\end{document}